# Tailoring Heat Transfer at Silica–Water Interfaces via Hydroxyl and Methyl Surface Groups

Viktor Mandrolko[1*], Konstantinos Termentzidis[2], David Lacroix[1] and Mykola Isaiev[1*]

[1]*Université de Lorraine, CNRS, LEMTA, F 54000 Nancy – France*

[2]*Univ. Lyon, INSA-Lyon, CETHIL UMR5008, F-69621, Villeurbanne, France*

Corresponding authors: Viktor Mandrolko, email: viktor.mandrolko@univ-lorraine.fr; Mykola Isaiev, email: mykola.isaiev@univ-lorraine.fr

## Abstract

Efficient thermal transport across solid-liquid interfaces is critical for optimizing heat dissipation in modern technological applications. This study aims to investigate how surface functionalization affects heat transfer at the silica/water interface, with specific objectives: (i) to assess the impact of surface modification on interfacial thermal resistance (ITR) and (ii) to elucidate the physical mechanisms governing heat transfer on functionalized surfaces. We employ non-equilibrium molecular dynamics (NEMD) simulations of silica surfaces with varying concentrations of methyl and hydroxyl groups to quantify ITR and analyze the contribution of each functional group to the total heat flux. Results demonstrate that transitioning from methylated to hydroxylated groups leads to: (i) an approximately sixfold increase in adhesion energy, (ii) a reorientation of interfacial water molecules perpendicular to the surface normal, (iii) a reduction in liquid depletion length near the interface, (iv) a nonlinear decrease of ITR correlated with an increase in the number and strength of interfacial hydrogen bonds, and (v) enhanced dynamic stability of OH-mediated bonds, with OH-donor lifetimes exceeding those of $CH_3$-donor bonds by a factor of 5–6 and showing peak persistence at intermediate hydroxylation levels (~50%). These findings highlight that manipulating the concentrations of functional groups enables precise tailoring of interfacial thermal transport, offering new opportunities for optimizing heat transfer in silica-based systems.

Keywords: Solid-liquid interface; Silica-water interface; Surface functionalization; Interfacial heat transfer; Kapitza resistance; Hydrogen bonding; Molecular Dynamics

# Introduction

Efficient heat management is a critical challenge in nanoscale systems, where thermal transport across solid–liquid interfaces can strongly influence system performance. At these scales, interfacial thermal resistance (ITR) can dominate over bulk conduction, dictating overall heat flux in applications such as microelectronics cooling [1], bio-nano interfaces [2] or electrochemical energy storage systems [3,4].

Silica ($SiO_2$) is one of the most abundant and versatile materials on Earth, with α-quartz being a common crystalline polymorph [5]. Its high thermal and chemical stability, mechanical strength [6], and, in crystalline form, piezoelectric properties [7] make it widely used in electronics and environmental applications [8-13]. Notably, while crystalline silica exhibits greater thermal resistance at the water interface compared to its amorphous counterpart [14,15], it also possesses higher bulk thermal conductivity[16], making it an attractive candidate for thermal management systems. Beyond its bulk properties, the silica surface is particularly intriguing, as it can be deliberately modified to tailor interfacial interactions and functionalities. Numerous studies focus on the silica-liquid interface [17-19], including silica surface structuring [20] and its natural hydroxylation [21]. It was demonstrated that the wettability and thermal resistance of silica-liquid interfaces depends predominantly on surface chemistry [15,22,23], opening the possibility of surface engineering to tune interfacial heat transport. For instance, introducing methyl groups improves hydrophobicity, while hydroxyl, amino, or carboxyl groups enhance hydrophilicity or reactivity [24–27].

Functionalization of silica surfaces is commonly achieved by reacting surface silanols with organosilanes [25–27]. There are multiple approaches such as sol-gel process, post-synthesis silylation, [24,25], one-pot or hybrid approaches which allow for the fabrication of surfaces with controlled chemical composition [24–27]. Methylation, in particular, converts hydrophilic Si–OH groups into hydrophobic $Si–CH_3$, increasing water contact angles above 140° [26,27] and reducing the work of adhesion, as described by the equation:

$$W_{adh} = \gamma_l \left(1 + \cos(\theta)\right) \qquad (1)$$

Here $\gamma_l$ is the liquid-vapor surface tension and θ is the apparent contact angle.

Experimental studies report a broad range of accessible hydroxyl densities on silica surfaces, from about 0.5 up to 9.8 –OH/nm$^2$ depending on morphology and pretreatment [28–30]. Subsequent surface methylation can replace a significant fraction of these groups, achieving different $CH_3$ coverage depending on the method [26,31–34], although complete substitution is typically hindered by steric limitations. These findings confirm the feasibility of partial to near-complete surface functionalization, enabling control over surface chemistry and expanding silica's applications in catalysis [35], drug delivery and biomedical devices [36–39], as well as environmental remediation and wastewater treatment [40–42]. Because surface chemistry directly affects the efficiency of heat transfer, this connection is often discussed in terms of wettability, and its correlation with interfacial thermal conductance.

While contact angle calculations are widely used to assess wettability, they become unreliable for highly hydrophilic surfaces where droplets spread into thin films, making the work of adhesion a more informative metric [43]. The relationship between wettability, adhesion, and interfacial thermal transport is complex and often system-dependent rather than universally consistent. In general, stronger wettability and adhesion enhance interfacial thermal conductance by promoting more efficient energy exchange across the solid–liquid boundary, and some studies indeed report a nearly linear dependence of conductance on the wetting angle [44–46]. However, this trend lacks universality: Acharya et al. [46] observed that the correlation between conductance and cos(θ) varies across systems, while Park and Cahill [47] demonstrated that solute effects can significantly alter conductance at hydrophobic Au/water interfaces. These and other findings indicate that wettability or adhesion alone is insufficient to predict thermal transport [46,48–51]. To address this variability, researchers have proposed the depletion length – the characteristic region near the interface where liquid structure rearranges due to surface interactions, as a more consistent descriptor. Ramos-Alvarado et al. [52] showed that depletion length correlates consistently with thermal conductance across smooth, non-polar surfaces like silicon and graphene-coated silicon, though coefficients vary for different materials (e.g., Pd, graphene oxide) [53,54]. For functionalized surfaces, additional mechanisms such as hybridized vibrational modes, electron-phonon coupling, and polaritonic interactions further complicate thermal transport predictions, particularly for polar molecule-functionalized surfaces [55,56].

To explore these relationships at the atomic level, molecular dynamics (MD) simulations have been extensively used to investigate the interfacial properties of silica and water [17,18,57–59]. For instance, Deng et al. [17] and Abramov et al. [58] showed that the (001) α-quartz surface is more hydrophilic than the (100) face due to its higher silanol density, while surface methylation reduces wettability. Bistafa et al.[18] analyzed the work of adhesion and interfacial entropy of hydroxylated silica, emphasizing their role in wetting thermodynamics. Bonnaud et al. [57] studied water confined in nanoporous silica, demonstrating that confinement alters both structural ordering and dynamics. Gonçalves and Termentzidis [22] further reported that grafting hydrophobic trimethylsilane on amorphous silica linearly decreases cos (θ), disrupts the hydrogen-bond network in interfacial water, and exponentially increases interfacial thermal resistance. Similarly, Goicochea et al.[15] investigated how modifying surface hydrophilicity and engineering surface nanostructure (e.g., nanopillars) on silica impact the global interfacial thermal conductance, showing that both chemical and topographical functionalization can boost heat transfer by enhancing vibrational coupling at the interface. Schoen et al. [60] directly compared interfacial conductance across hydrophobic and hydrophilic silica-water interfaces, demonstrating that thermal transport is significantly higher for hydrophilic (silanol-terminated) surfaces, and that this enhancement is closely linked to the presence of robust hydrogen-bond networks.

While these studies collectively establish the importance of surface chemistry in determining both wettability and interfacial heat transfer, most of them treat the interface as a homogeneous entity and evaluate only the global ITR. The specific contributions of distinct surface groups, such as silanols (–OH) and methyls (–$CH_3$), to the overall heat flux remain poorly resolved. To address this gap, the present study explicitly disentangles the role of individual functional groups and, additionally, investigates the hydrogen bond lifetimes formed between interfacial water and each group type, linking local bonding dynamics to interfacial thermal transport.

The following sections describe the employed MD framework and force field parameters, followed by results that connect surface functionalization to both wetting and heat transfer phenomena. Key interfacial metrics: contact angle, work of adhesion, ITR, depletion length, and water orientation are evaluated to establish quantitative links between surface chemistry and thermal behavior. The article concludes with a discussion of these findings and their broader implications for the design of silica-based systems.

# Materials and Methods

For simulations, classical molecular dynamics was employed, implemented through the LAMMPS package [61,62]. Silica was simulated with the modification of alpha-quartz (001) surfaces functionalized with hydroxyl (-OH) and methyl (-CH3) groups. The initial atomic coordinates of the pristine alpha-quartz slab were generated using the Materials Project database[63], while individual structures of the methyl (-CH3) and hydroxyl (-OH) groups were taken from the PubChem database [64]. Functionalization was performed by covalently grafting either hydroxyl or methyl groups to the silicon atoms located on the slab surface using a custom Python script. The grafting was performed by replacing the two terminal oxygen atoms bonded to each surface Si atom with either two hydroxyl or two methyl groups. Cartesian coordinates of -OH and $-CH_3$ groups were oriented such that their bond direction radiated outward from the surface along the vector connecting the surface Si atom and the corresponding terminal O atom. The density of functionalization groups reached the maximum possible value (9.135 groups/$nm^2$), which results from functionalization of each surface Si atom with either 2 hydroxyl or 2 methyl groups. Scripts used for generating these initial functionalized systems are publicly available on GitHub[65] to ensure transparency and reproducibility.

To simulate different statistical arrangements, three independent seeds were created for each functionalization variant. In each seed, the atoms selected for functionalization were chosen randomly. All modifications and coordinate manipulations were performed with custom Python scripts. All input structure files, scripts, and random seeds are available from the corresponding author upon request.

We utilized the ClayFF force field [66], known for its comprehensive treatment of bonded and non-bonded interactions in clay minerals and related oxide systems[67]. ClayFF was originally parametrized using structural and energetic data from quantum mechanical calculations (DFT) and experimental measurements of clay minerals, oxides, and hydroxides and has been widely validated for simulating silica phases, including quartz and amorphous silica [58,68-71]. Comparative studies have shown that ClayFF provides close agreement with ab initio MD and X-ray reflectivity data for quartz–water interfaces, accurately reproducing the interfacial water structure and hydrogen-bonding patterns [71,72].

For the surface methyl groups (Si-$(CH_3)_2$), parameters were adapted from the OPLS-AA force field [73] and previous studies[69,74-76], as detailed in Deng Y et al.[17]. Although the ClayFF force field has been reported to overestimate electrostatic interactions on bare silica surfaces, resulting in overly hydrophilic behavior, it performs reliably for functionalized surfaces[77], which are the focus of this work. In these systems, the presence of –OH and –$CH_3$ groups screens the surface charges that drive excessive polarization of water, moderating the interaction strength. This limitation may lead to slightly lower absolute ITR values for the most hydrophilic configurations, since stronger water–surface coupling facilitates heat transfer, but it does not affect the relative trends discussed in this study. In ClayFF, bonded interactions for functional groups are defined by harmonic potentials for bond stretching and angle bending, while non-bonded interactions are governed by 12 – 6 Lennard-Jones and Coulomb potentials. Water was simulated in frames of SPC/E model [78], chosen for its simplicity and compatibility with ClayFF [67]. Since all atoms carry partial charges, Coulombic interactions occur between all pairs of atom species. For Van der Waals interactions, hydrogen atoms in water molecules and hydroxyl groups are excluded, consistent with the standard ClayFF and SPC/E formulations [17,67]. Detailed interaction parameters are provided in the Supporting Materials (SM).

Several silica surface functionalization variants were investigated, ranging from complete hydroxylation to complete methylation. The fully hydroxylated surface corresponds to $N_{OH}/N \cdot 100\% = 100\%$, where $N$ is the total number of functional groups, and $N_{OH}$ is the number of hydroxyl groups. Intermediate functionalization states were also considered, with decreasing proportions of hydroxyl groups randomly replaced by methyl groups: $N_{OH}/N \cdot 100\% = 75, 50, 37.5, 25, 12.5\%$. The opposite extreme, where all functional groups are methyl, corresponds to $N_{OH}/N \cdot 100\% = 0\%$. It is worth emphasizing that we considered a fully functionalized surface in each case because the pure silica surface is highly reactive [79]. All simulations, including those for contact angle, adhesion, and ITR were performed with a time step of 0.5 fs.

## Wetting angle calculation

In order to calculate wetting angles, we constructed a system comprising a functionalized silica substrate supporting a droplet in its initial state. By implementing periodic boundary conditions on two sides of the system, we obtained a cylindrical droplet. This choice aimed to simplify the spatial variation of contact angle along the three-phase contact line of a hemispherical droplet and minimize the influence of droplet size effects [80,81]. The right panels of Figure 1a show the surface functionalization states before and after relaxation and thermalization for three functionalization variants: fully methylated, 1:1 methyl/hydroxyl, fully hydroxylated. Prior to constructing the surface models, the bulk α-quartz structure was relaxed through energy minimization with simultaneous cell and atomic position relaxation under zero external pressure to eliminate initial structural stresses and achieve a mechanically equilibrated state before dynamics. The minimization was performed using the conjugate gradient algorithm, targeting an energy tolerance of $1.0\times10^{-5}$ eV and a force tolerance of $1.0\times10^{-7}$ eV/Å, with a maximum of 100,000 iterations and 400,000 force evaluations. During minimization, the box dimensions were allowed to change isotropically to ensure mechanical equilibrium at 0 bar pressure. The final box dimensions deviated by less than 2% from their initial values, confirming that the constructed system was already close to equilibrium before molecular dynamics runs.

The initial, unrelaxed configurations are shown alongside the relaxed ones to illustrate how the surface structures evolve during equilibration. From the snapshots of thermalized surfaces, it is seen how the spatial arrangement of surface groups varies depending on their nature. In the fully methylated surface, the relatively large $CH_3$ groups occupy significant space around each silicon atom, which leads to one of the two groups being oriented toward the interior of the material, while the other points outward. In contrast, on the fully hydroxylated surface, the smaller OH groups tend to bend toward the surface, forming a dense network of hydrogen bonds (see the right part of Figure 1a). The bottommost 2 Å of the substrate atoms were restrained to their initial positions using spring/self with a harmonic spring constant of 0.01 eV/Å$^2$ in the $z$ direction to mimic the bulk constraints. The value was chosen empirically: we tested a range of spring constants and selected 0.01 eV/Å$^2$ as a compromise that is sufficiently soft to allow relevant low-frequency substrate vibrational modes to develop, while still preventing significant translational drift of the constrained atoms.

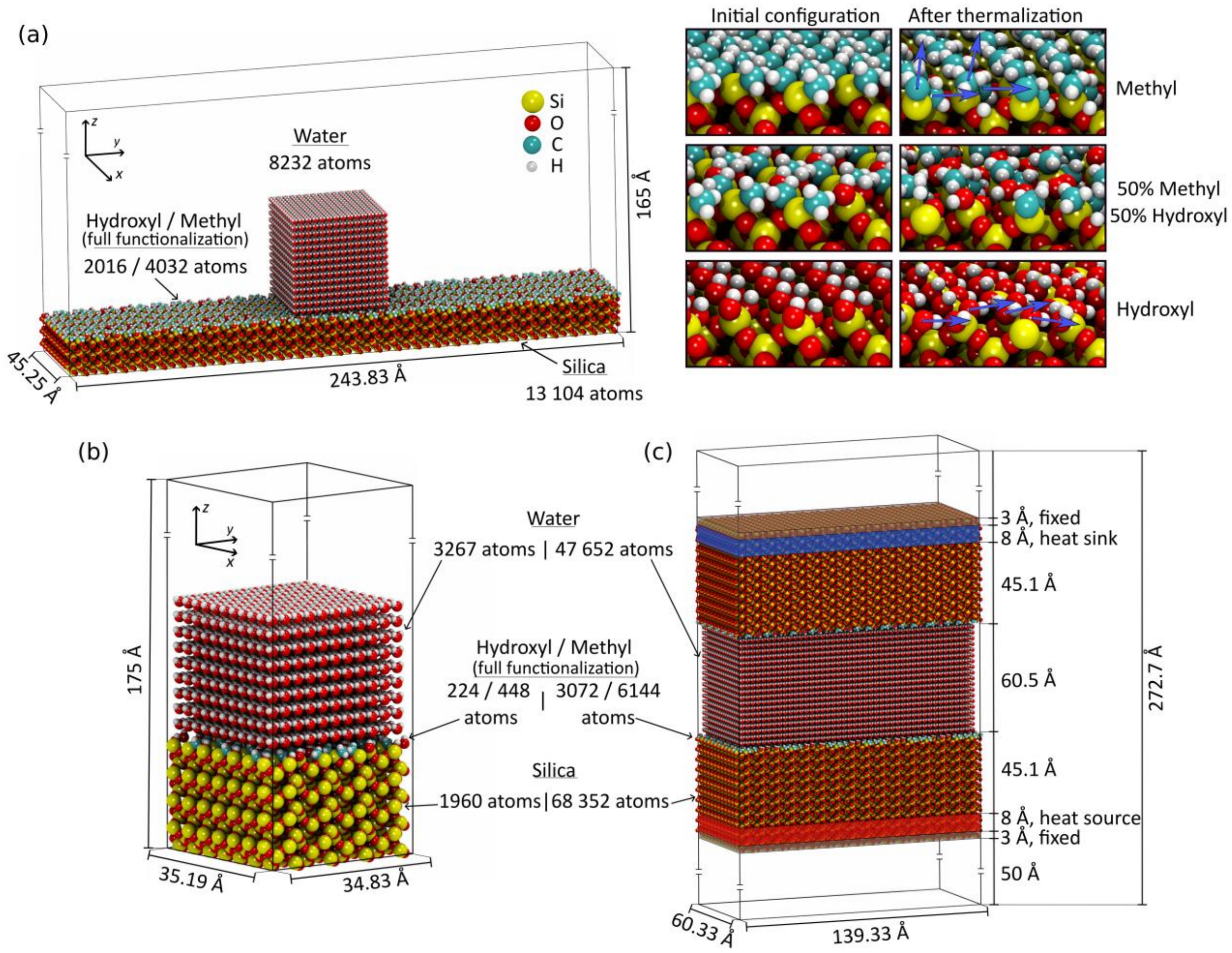

*Figure 1. (a) Initial configuration of a water droplet on a functionalized silica surface (left) and representative surface snapshots before and after thermalization for three functionalization variants: fully methylated, mixed methyl/hydroxyl (1:1), and fully hydroxylated (right). (b) Model system used for the work of adhesion calculation. (c) Model system employed in the thermal transport simulations.*

The dimensions of the simulation box and the quantities of atoms for each species (Si , -$CH_3$, -OH, and $H_2O$) are detailed in Table 1.

Table 1. Characteristic lengths of the simulation box and the number of atoms of each species for the system used for the wetting angle calculation.

| | | |
|---|---|---|
| Simulation box | | 45.25 Å×243.83 Å×165.0 Å |
| Silica substrate | | 45.25 Å×243.83 Å×17.40 Å |
| Initial configuration of a water droplet | | 40.3 Å×40.3 Å×40.3Å |
| Number of atoms | Silica substrate | 13 104 |
| | Water | 8 232 |
| | Methyl (-$CH_3$) (fully functionalized) | 4032 |
| | Hydroxyl (-OH) (fully functionalized) | 2016 |

The number of water molecules was selected to produce a droplet size commonly used in molecular simulations of wetting [81-83]. This ensures a balance between computational efficiency and physical accuracy: smaller droplets are more susceptible to curvature-dependent deviations in surface tension (Tolman correction) [80,84], which would in turn affect the measured contact angle.

The simulation was performed in the following order: first, the entire system, including both the water droplet and the silica substrate, was gradually heated from 1 to 300 K in four stages using temperature rescaling, followed by 1.2 ns of thermalization at 300 K in the NVT ensemble. Next, it was performed in the NVT ensemble for 10 ns. The pressure was not explicitly constrained, as well as in all other systems simulated in this study. This choice was motivated by the fact that local pressure fluctuations in nanoscale systems reach hundreds of atmospheres, largely exceeding the effect of moderate external pressure variations. Therefore, applying an external barostat was considered unnecessary. During this time, 20 density profiles were acquired. The coordinate data were recorded every 0.5 ps, and each density profile was constructed by averaging over 1000 consecutive snapshots, corresponding to 0.5 ns of simulation time. The bin size used for computing the density profiles along the z-direction was 0.5 Å. The wetting angle was found for each density profile by approximating the droplet surface with a circle. To avoid distortions from the interfacial entropy-affected zone near the solid surface (i.e., layering and structuring effects), we excluded a thin region of the droplet within the depletion length above the substrate from the circle fitting process, considering only density isosurfaces beyond this region for the geometrical fitting.

Next, the contact angle was plotted as a function of time, and the average value was found. Only the wetting angles found in the last 5 ns were considered to ensure the droplet stabilized and reached the optimum contact angle. To ensure statistical reliability and account for variability due to initial conditions, each functionalization variant was simulated three times with different random distributions of functional groups and randomized atomic velocities. The reported contact angles represent averaged values over these independent simulations, and standard deviations are provided as error bars in Figure 7.

## Work of adhesion calculation

The work of adhesion $W_{adh}$ was calculated using the "phantom wall" method [85]. This approach introduces a repulsive potential that selectively interacts with water molecules, allowing controlled liquid separation from the surface. By gradually displacing the wall and measuring the corresponding forces, one can determine the energy required to detach the liquid from the substrate, which directly corresponds to the work of adhesion. This method is widely used due to its conceptual simplicity and robustness [18,86-88]. Importantly, its derivation relies only on the reversible mechanical separation of the liquid using a parametrized external potential and the evaluation of the corresponding forces along the integration path. As a result, it is inherently compatible with any interatomic potential that provides force information. In practice, the interactions between the solid substrate and the liquid are described by their respective interatomic potentials, while the interaction between the phantom wall and the liquid is defined by a simple purely repulsive LJ-type potential. The wall acts only as a virtual mechanical constraint and does not affect the underlying solid–liquid potential energy surface. This method was validated in water-silica systems with bonded force fields [18,89], yielding $W_{adh}$ values consistent with experiments and alternative methods. Beyond these water–silica benchmarks, the phantom-wall thermodynamic-integration scheme has also been shown to reproduce the same qualitative dependence of solid–liquid interfacial free energies and wetting transitions as mechanical (stress–anisotropy) methods applied to LJ model systems [90,91]. Unlike the mechanical route, the phantom-wall method does not depend on any microscopic definition of the stress tensor and is not affected by residual stresses or finite-size effects in the solid slab, which makes it more robust for solid–liquid interfaces with atomistic solids [92].

We considered a system comprising a solid functionalized silica slab, fully covered by a water layer with a height three times greater than the Lennard-Jones cut-off distance. The characteristic sizes of the system are presented in Figure 1b (in Table 2.) For the adhesion calculations, the bottommost 2 Å of the silica slab were restrained using the same spring/self protocol as in the contact-angle systems (harmonic spring constant of 0.01 eV/Å$^2$ in the z direction), to mimic the mechanical constraint of the bulk substrate while preserving realistic interfacial dynamics.

Table 2. Characteristic lengths of the simulation box and the number of atoms of each species for the system used for the work of adhesion calculation.

| | | |
|---|---|---|
| Simulation box | | 35.19 Å×34.83 Å×175.39 Å |
| Silica substrate | | 35.19 Å×34.83 Å×22.98 Å |
| Initial configuration of a water droplet | | 31.0 Å×31.0 Å×24.8Å |
| Number of atoms | Silica substrate | 1960 |
| | Water | 3267 |
| | Methyl (-$CH_3$) (fully functionalized) | 224 |
| | Hydroxyl (-OH) (fully functionalized) | 448 |

The system was first heated and maintained at 300K during 0.5 ns. A repulsive phantom wall (i.e., a virtual boundary) perpendicular to the z-direction was then introduced at $z =$ 18.5 Å, below the interface and beyond the wall–water interaction cut-off distance, to ensure it did not interact with the water molecules. The wall potential was set to interact exclusively with water molecules (oxygen and hydrogen atoms) and to be transparent for silica. The forces exerted by the wall were computed using the standard LAMMPS command *fix wall/lj126*, which applies a 12–6 Lennard-Jones potential with parameters $\epsilon = 6.9382$ Å, $\sigma = 3.16$ Å and cut-off distance 3.546 Å, rendering the wall repulsive only. The choice of potential is not critical, as long as it ensures a purely repulsive interaction with water molecules and allows the force exerted by the wall to be computed reliably. The wall was then gradually raised to a distance of 9 Å (up to $z = 27.5$ Å) in 70 uniform spatial steps. At each spatial step, the system was first thermalized for 0.1 ns, and then the value of the force exerted by the wall on the water and the system's energy were averaged over 0.5 ns. All calculations were carried out in the NVT ensemble.

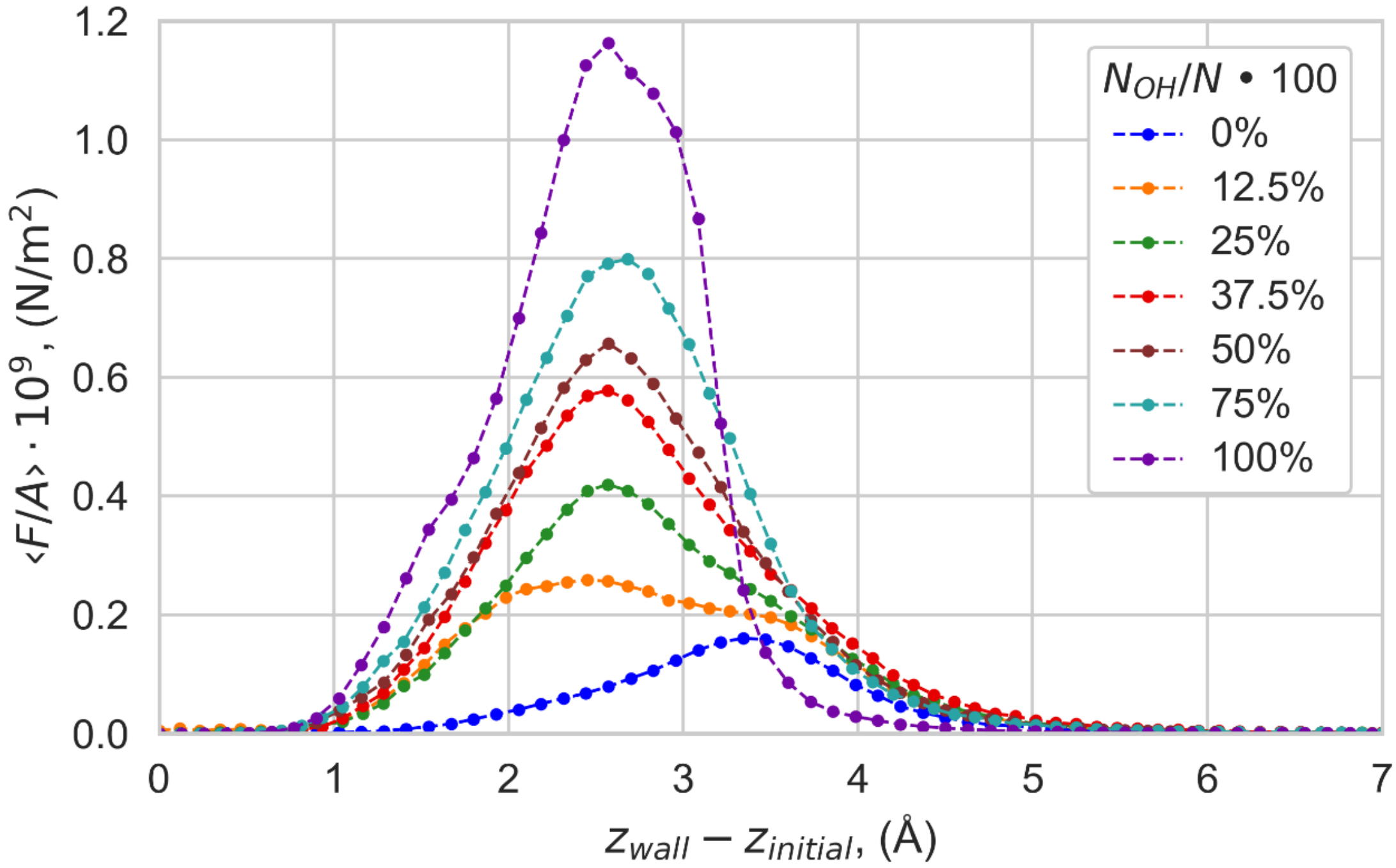

*Figure 2. Average force per unit area exerted by the phantom wall on water as a function of its displacement from the initial position. The phantom wall acts as a virtual boundary interacting exclusively with water molecules. Different curves represent varying ratios $N_{OH}/N$ of hydroxyl groups*

Figure 2 depicts the dependencies of the average force per unit area exerted by the phantom wall on water as a function of its displacement from the initial position. To compute the work of adhesion from this force profile, we adopt the thermodynamic integration approach. A coupling parameter $\lambda \in [0,1]$ is introduced to quasi-statically move the wall from its reference position $z_{initial}$ to the target position $z_{final}$ via:

$$z_{wall} = \lambda(z_{final} - z_{initial}) + z_{initial} \qquad (2)$$

This defines a reversible path in the system's Hamiltonian $H(\lambda)$ , connecting the initial $(\lambda = 0)$ and final $(\lambda = 1)$ equilibrium states. The corresponding free energy change is given by the standard thermodynamic integration expression[93]:

$$\Delta F = \int_0^1 \langle {}^{\partial H}/_{\partial \lambda} \rangle d\lambda \qquad (3)$$

Where $\partial H/_{\partial\lambda}$ is the force $F(z)$ exerted by the wall on the fluid scaled by $(z_{final} - z_{initial})$. Changing variables from $\lambda$ to z and dividing by the interfacial area $A$, one obtains work of adhesion:

$$W_{adh} = \frac{\int_{z_{initial}}^{z_{final}} F(z)dz}{A} \quad (4)$$

which we compute numerically from the discrete data shown in Figure 2.

At the same time, the molecular dynamics method provides us with the capability to directly observe the change in the potential energy of the interaction between the solid and the liquid $\Delta U$ , which allows us to measure the interfacial entropy change per unit area $T\Delta S$ [94]:

$$T\Delta S = \Delta U - W_{adh} \quad (5)$$

The expression arises from decomposing the interfacial free energy change into enthalpic and entropic contributions. As demonstrated previously [94,95], liquid-liquid interactions are thermodynamically compensating and do not impact the net free energy. Thus, the work of adhesion depends solely on solid-liquid interactions, expressed as $W_{\mathrm{adh}} = \Delta U - T\Delta S$, where $\Delta U$ and $T\Delta S$ represent the potential energy increase and entropy gain during detachment, respectively. Here, $\Delta U$ is positive due to the energy input required to overcome attractive forces, while $T\Delta S$ also increases as the system gains disorder upon separation.

## Interfacial thermal resistance

The non-equilibrium molecular dynamics method (NEMD) was employed to establish the relation between the functionalization of silica surface and the interfacial thermal transport. Particularly, the heat flux through the interface and ITR were evaluated. The studied system is depicted in the Figure 1c. Similar to Klochko et al.[45], the simulations were performed with periodic boundary conditions, while sufficient empty space was left in $z$ direction to prevent contact between the top and bottom silica slabs. We considered the following functionalization variants: $N_{OH}/N \cdot 100\% = 0, 10, 30, 50, 70, 90, 100\%$. The simulation procedure was as follows: first, the system was heated and thermalized at 300 K for 0.35 ns in the NVT ensemble. During relaxation, the liquid volume changed and reached different equilibrium values depending on the surface functionalization.

Nevertheless, the water region remained sufficiently large to ensure the presence of a bulk region in the center. We present the volume variation, as well as the change in bulk water density, in Figure 3.

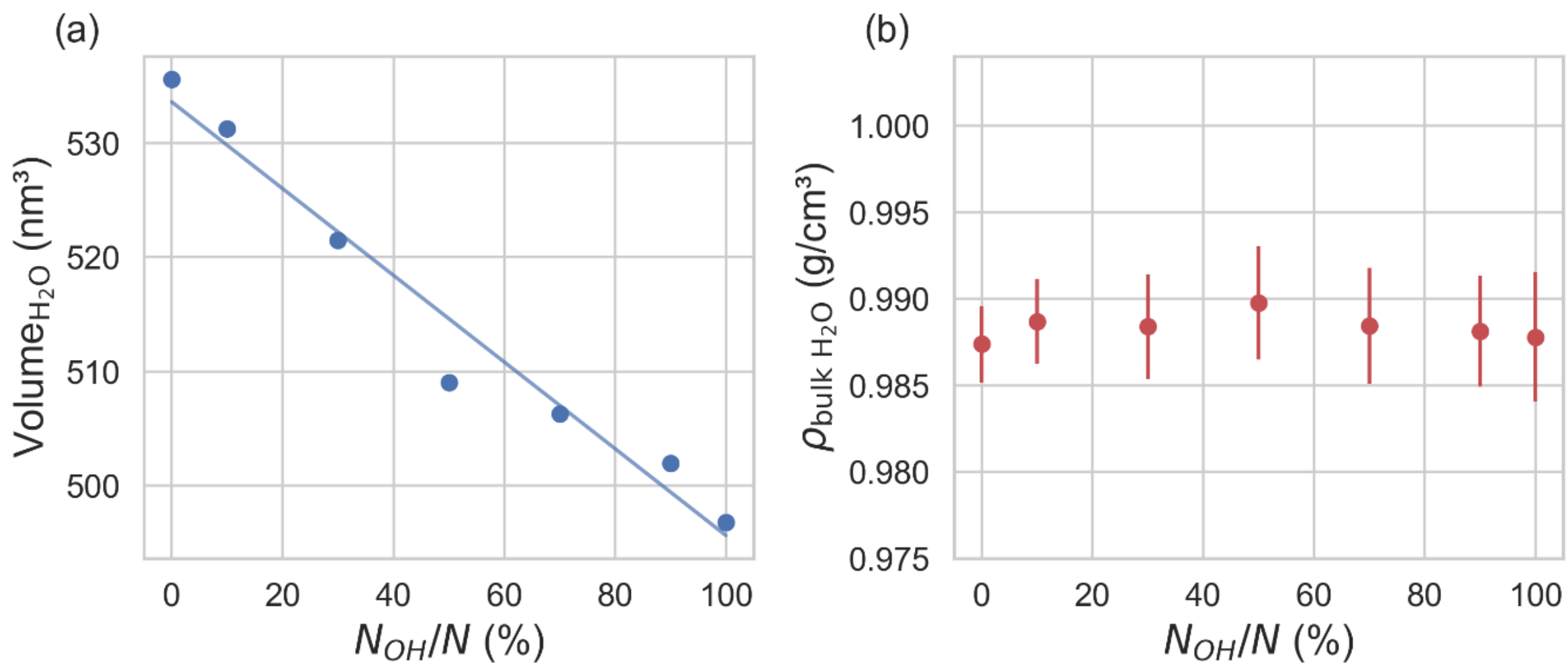

*Figure 3. Dependence on the degree of surface functionalization of (a) the volume of the liquid phase and (b) the density of bulk water. The blue line serves as a linear approximation. The bulk region was defined as a 30 Å thick slab located at the center of the liquid domain. Error bars represent the standard deviation of the average density.*

The liquid domain was defined as the region between two equimolar surfaces, calculated using Equation ( 10 ). As shown in Figure 3, the total volume of the water phase decreases with increasing surface density of hydroxyl groups. This trend is expected, as stronger interactions between the polar OH groups and water molecules lead to a more compact and structured interfacial layer, resulting in an overall reduction of the system's liquid volume. Despite this volume reduction, the central region of the water domain remained large enough to preserve bulk-like behaviour. The density of bulk water, measured in a 30 Å thick slab at the centre of the domain, remained nearly constant across all functionalization variants, with deviations falling within the margin of statistical uncertainty. This confirms that the presence of a stable bulk region was maintained throughout the simulations.

After the thermalization, the most top and bottom layers of atoms with a width of 3 Å were tethered in the *z* direction using the spring/self restraint protocol established for the contact-angle systems (k = 0.01 eV/Å$^{-2}$). The heat source and sink (8 Å each) were implemented by rescaling atomic velocities in the corresponding regions to maintain temperatures of 330 K and 270 K, respectively, inducing heat flow through the system. After, the system was held in NVE ensemble for 0.25 ns to reach the equilibrium heat flow, which means that the amount of heat generated by the source is equal to the amount of heat dissipated by the sink. Subsequent data accumulation was performed over 1 ns during which average amount of energy passing through the system as well as the one-dimensional density profile was obtained. To verify the reliability of the ITR values, each functionalization variant was simulated at least three times with independently generated functional group distribution and initial atomic velocities. The resulting thermal resistance values were found to be consistent across repetitions, confirming the reproducibility of the results. In addition, convergence tests were carried out (see SM, Figure S1) and confirmed the temporal stability of both the temperature profiles and the corresponding ITR values.

Given the total amount of energy $\langle E \rangle$ passing through area $A$ in time $t$, the heat flux was defined as:

$$J = \frac{\langle E \rangle}{A \cdot t} \tag{6}$$

In addition to the heat flux calculation method described above, we also employed an alternative approach that enables the decomposition of heat flux based on the generalized Irving–Kirkwood formalism[96], which enables decomposition of the heat flux into contributions from different atom groups. The intergroup heat flux is given by:

$$J_{group\ 1 \rightarrow group\ 2} = \sum_{i \in group\ 1} e_i\, \vec{v}_i + \frac{1}{2} \sum_{i \in group\ 1} \sum_{\substack{j \in group\ 2 \\ j \neq i}} \left( \vec{F}_{ij} \cdot \vec{v}_j \right)\, \vec{r}_{ij} \tag{7}$$

Here, $e_i$ and $\vec{v}_i$ are the per-atom energy and velocity, respectively; $\vec{F}_{ij}$ is the force exerted by atom $i$ on atom $j$; and $\vec{r}_{ij}$ is the distance vector between them. The formula evaluates the heat flow within the first group, consisting of a convective part from atomic motion and a virial part resulting from interactions with atoms in the second group.

In the present work, we used the standard LAMMPS implementation of this expression via the *compute heat/flux/virial/tally* command. This compute evaluates the virial component of the heat flow between two predefined atom groups without explicitly recalculating forces or energies, using internal pairwise-force callbacks within the LAMMPS kernel. Although the convective contribution is not included in this particular implementation, it becomes negligible for solid–liquid interfaces, where no mass transport occurs across the boundary. In this case, the virial term alone correctly represents the total interfacial heat transfer, consistent with the method of planes formulation. In practice, we defined multiple solid and liquid atomic groups (substrate, bulk Si, O, surface OH, and $CH_3$ groups) and evaluated their pairwise energy transfer rates to the liquid water phase. The resulting instantaneous heat flow rates were then averaged in time using the *fix ave/time* command to obtain statistically converged interfacial fluxes for both the upper and lower interfaces. The total interfacial heat flux was then obtained by averaging the results over both interfaces to account for symmetry and reduce statistical fluctuations.

To evaluate the interfacial thermal resistance, we first computed the steady-state one-dimensional temperature profile along the $z$-axis using binning with a spatial resolution of 0.5 Å. The temperature in each bin was calculated based on the kinetic energy according to the equipartition theorem. For rigid water molecules constrained by the SHAKE algorithm, three out of nine intramolecular degrees of freedom are frozen, resulting in only two-thirds of the actual temperature being reported by LAMMPS. To correct for this, the binned water temperature values were multiplied by a factor of 1.5. For all other atoms, the default temperature evaluation provided by LAMMPS was used without modification. The total temperature profile was calculated from temperature profiles of the solid and liquid regions as density-weighted averages:

$$\langle T \rangle = \frac{\rho_{solid} T_{solid} + \rho_{liquid} T_{liquid}}{\rho_{solid} \rho_{liquid}} \tag{8}$$

The temperature jump $\Delta T$ was then obtained as the difference between the extrapolated linear fits of the solid and liquid temperature curves at the interface position $z_{eq}$. The example of temperature distribution with temperature jumps is shown in Figure 4:

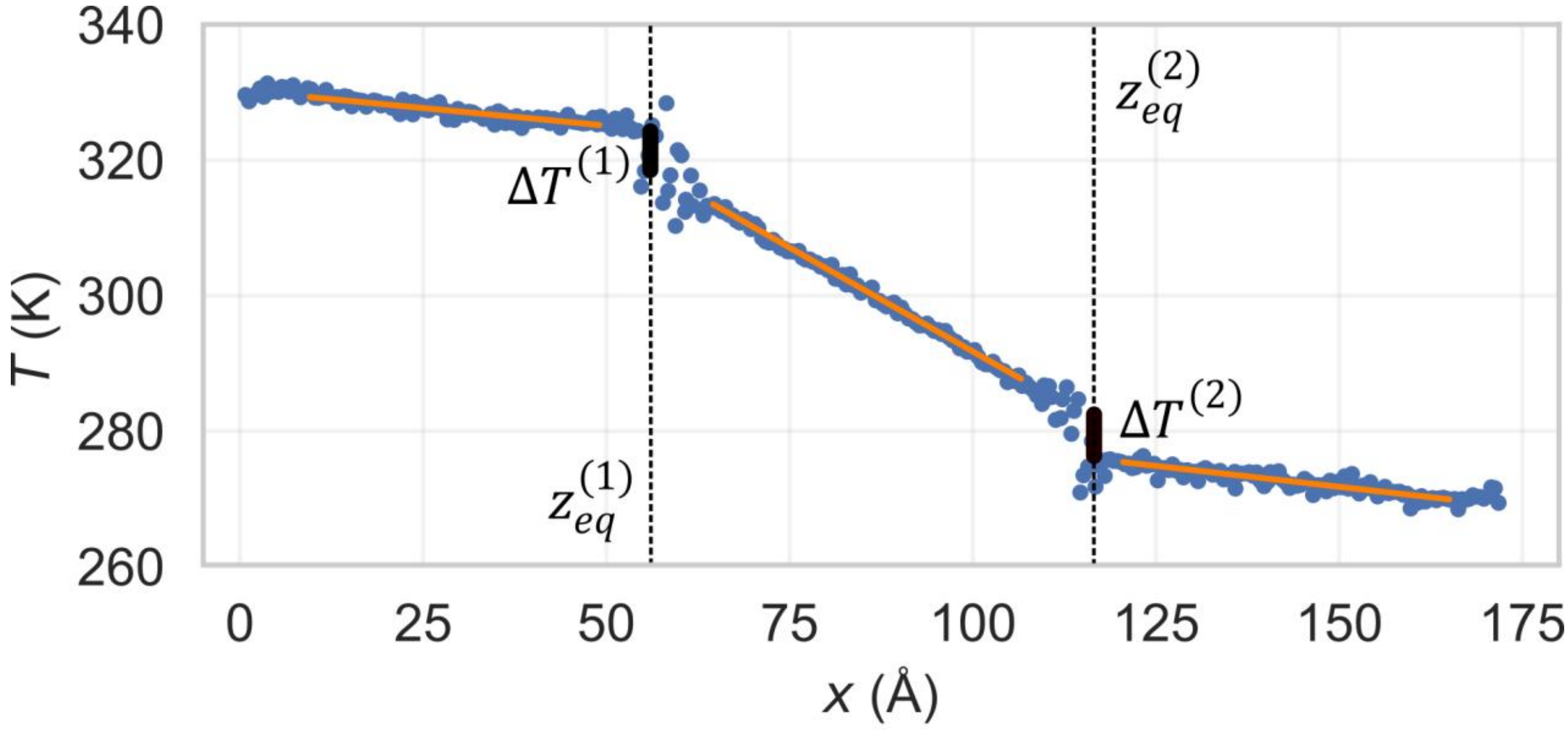

*Figure 4. Example of a temperature profile for a system with the functionalization variant $N_{OH}/N \cdot 100\%$ = 50%. The dashed vertical lines indicate the equilibrium positions of the solid–liquid interfaces. The temperature jumps at the interfaces are highlighted in bold black. The final temperature jump is taken as the average of the two jumps measured at both solid–liquid boundaries.*

To accurately determine $z_{eq}$, we used the density profile $\rho(z)$ and applied the concept of the Gibbs dividing surface [43,97], which defines the interface as the position that balances the excess density on both sides. This condition is expressed as:

$$\int_{-\infty}^{z_{eq}} (\rho_1 - \rho(z))dz = \int_{z_{eq}}^{\infty} (\rho(z) - \rho_2)dz \qquad (9)$$

where $\rho_1$ and $\rho_2$ are the bulk densities of both phases. Equation ( 9 ) represents the theoretical definition of the Gibbs dividing surface in a continuous system, ensuring that the excess density deficit on one side of the interface equals the excess density surplus on the other. However, in molecular dynamics simulations, the density profile is not continuous but is instead obtained as discrete data points due to the binning process (with a spatial resolution of 0.5 Å in this case). Therefore, we approximate the interface position by considering a finite layer around the interface where the density transitions from $\rho_1$ to $\rho_2$. This layer is defined with a starting coordinate $z_0$ and a width $d$, and we compute the average density within this layer, denoted $\langle\rho\rangle$. This leads to the practical formula:

$$z_{eq} = z_0 + \frac{\langle \rho \rangle - \rho_2}{\rho_1 - \rho_2} \cdot d \qquad (10)$$

To clarify how Equation ( 10 ) is derived from Equation ( 9 ), we provide the explanation in SM. Finally, knowing the heat flux $J$ and temperature jump $\Delta T$, the interfacial thermal resistance $R$ can be calculated as:

$$R = \frac{\Delta T}{J} \qquad (11)$$

## Water molecules orientation

Since the thermal transport across the water/solid interface may be connected with the water molecules' orientation[98], we calculated the probability of orientation of a water molecule relative to the z-axis based on 400 snapshots of the system that was used to calculate the thermal resistance. During the calculations, we considered a near-surface layer of water with a thickness of 10 Å, considering only one water/substrate interface, with the water located above the substrate along the z-axis.

For each water molecule in this layer, we calculated the azimuthal angle $\theta$, defined as the angle between the substrate-normal $z$-axis and the water molecule's dipole vector. Using these angles, we constructed a histogram of the orientation distribution with a bin width of $\Delta\theta$ = 1°, spanning from 0° to 180°.

However, in three-dimensional space, the volume of each angular bin depends on $\theta$ due to the spherical coordinate system. Specifically, the volume element is given by:

$$\Delta V \approx \frac{2\pi r^3}{3} \sin\theta \ \Delta\theta \qquad (12)$$

For a fixed radial distance and azimuthal symmetry, the probability density $P(\theta)$ must account for the geometric factor $\sin\theta$. To ensure uniformity, the raw count of molecules in each bin, $N_{H_2O}$ $(\theta)$, was normalized by $\sin\theta$ and the bin width:

$$P(\theta) = \frac{N_{H_2O}(\theta)}{\sin\theta \cdot \Delta\theta} \qquad (13)$$

Figure 5 shows the schematic of the 3D orientation angle $\theta$ and the azimuthal bin $\Delta\theta$ used to compute the orientation distribution of water molecules relative to the substrate-normal $z$-axis.

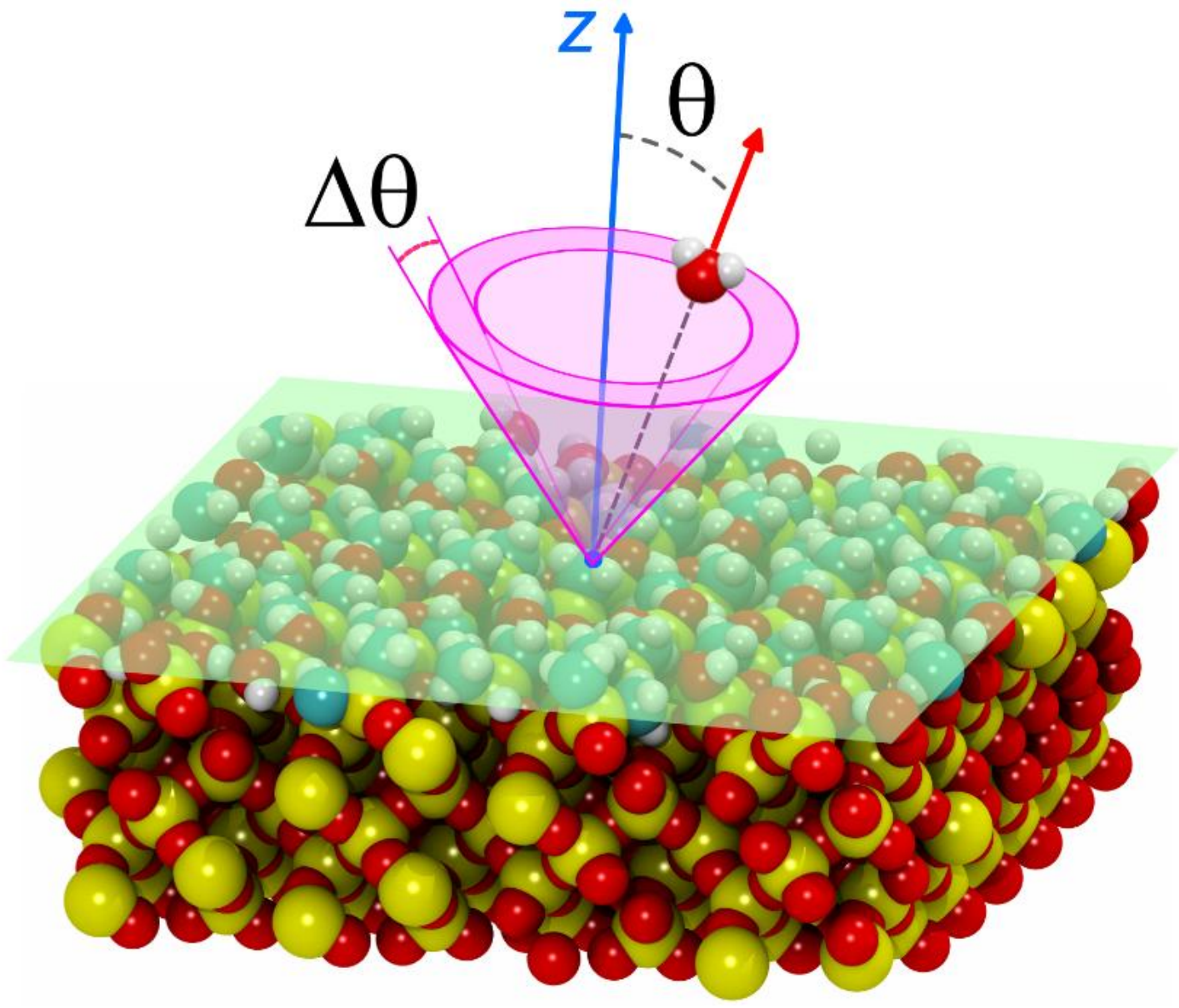

Figure 5. 3D schematic of water-molecule orientation: the blue axis is the substrate-normal $z$-axis, the red arrow shows the water dipole vector, and the magenta cone illustrates an angular bin of width $\Delta\theta$ = 1° at a fixed radial distance. The volume element scales with sin $\theta$ in spherical coordinates.

To quantify the fraction of water molecules with specific orientations, we defined the percentage of molecules whose orientation falls within a given angular interval $[\theta_a, \theta_b]$ as:

$$f_{[\theta_a,\theta_b]} = 100 \cdot \frac{\int_{\theta_a}^{\theta_b} P(\theta)\, d\theta}{\int_0^{180^\circ} P(\theta)\, d\theta} \qquad (14)$$

To complement the angular probability distributions, we further quantified the degree of orientational ordering using the standard nematic orientational order parameter [99]:

$$S = \frac{1}{2}(3\langle cos^2\theta\rangle - 1) \qquad (15)$$

which ranges from $S$=1 (perfect alignment along the surface normal), through $S$=0 (random/isotropic), to $S$=−1/2 (perfect alignment parallel to the interface). This parameter compresses the full orientational distribution into a single scalar quantity. To evaluate the order parameter locally near specific surface groups, we computed $S$ for water molecules whose oxygen atoms lie within a cutoff distance of 3 Å from the oxygen atoms of –OH groups and 4 Å from the carbon atoms of $–CH_3$ groups. The slightly larger cutoff for $–CH_3$ groups was necessary due to the weaker water $–CH_3$ interactions and the reduced local density of water at hydrophobic sites.

## Interfacial hydrogen bonding

To investigate the role of interfacial hydrogen bonding in thermal transport across the water/solid interface, we analyzed the average number of hydrogen bonds formed by functional groups on the substrate surface. The analysis was performed using atomic coordinates from xyz files generated for the systems used to calculate thermal resistance. These systems comprised 200 snapshots per simulation, with a time interval of 5 ps between snapshots, corresponding to a total simulation time of 1 ns per seed. For each level of surface functionalization, a minimum of three independent seeds were simulated, resulting in an ensemble average over 600 snapshots (equivalent to 3 ns of simulation time). Both the upper and lower substrates were considered to capture the hydrogen bonding behavior at both interfaces.

Hydrogen bonds were categorized into three types based on the functional groups on the surface: $CH_3$ donor (C-H…O), OH donor (O-H…O), and OH acceptor (O…H-O) H-bonds. Notably, there is no category for $CH_3$ acceptor (C…H-O) H-bonds. This is because the carbon atom in the methyl group lacks lone pairs of electrons, which are essential for acting as hydrogen bond acceptor[100]. Non-trivial C-H…O bonds, though weak, are experimentally validated in via NMR and quantum calculations [100-102]. To establish the existence of a hydrogen bond, the criteria proposed by Ozkanlar and Clark[103] were used: a hydrogen-acceptor distance $r_{H-A}$ < 2.5 Å and an angle $\alpha$ > 150°, defined as the angle between the donor-hydrogen bond vector (D-H) and the hydrogen-acceptor vector (H…A). These criteria were applied to all hydrogen bond types, where D represents the donor atom (O or C) and A is the acceptor atom (O).

To complement the static analysis of interfacial hydrogen bonding, we additionally quantified their dynamic behavior in terms of formation and rupture rates and average hydrogen bond lifetime. To this end, we performed new simulations of the same systems, but with higher temporal resolution: atomic coordinates were saved every 10 fs for a total duration of 100 ps (10,000 frames). This finely sampled trajectory enables tracking of individual hydrogen-bonding events over time.

For each frame, all interfacial hydrogen bonds were identified using the same geometric criteria. The hydrogen-bonding data were then stored in serialized Python objects (.pkl files) for efficient access. These pickle files contained, for each time frame, sets of unique hydrogen bonds classified by type, along with the corresponding atom indices.

Subsequently, another script was used to analyze the time evolution of these bonds and extract dynamic parameters. For each bond type, the number of hydrogen bonds was monitored as a function of time, and bond formation and rupture events were counted between consecutive frames. The formation rate was computed as the total number of newly formed bonds divided by the total simulated time, while the rupture rate corresponded to the number of broken bonds normalized in the same way. For transparency and reproducibility, all scripts used for hydrogen bond identification and analysis are publicly available on GitHub [65].

To further extend the investigation of the dynamics of H-bond, their lifetimes were quantified using both continuous and intermittent definitions of the H-bond autocorrelation function $C(t)$ [104,105], which measure the probability that a bond present at time $t_0$ still exists at time $t_0 + t$.

For the continuous case, the autocorrelation function was computed as:

$$C_c(t) = \frac{\langle H_c(0,t) \rangle}{\langle h(0) \rangle} \tag{16}$$

The function $H_c(0,t)$ equals 1 only if the hydrogen bond remains continuously present during the entire time interval $[0,t]$, and 0 otherwise. In practice, this means that once a bond is broken at any intermediate frame between 0 and $t$, $H_c(0,t)$ is set to 0 for that time origin. Therefore, the continuous $C_c(t)$ is particularly sensitive to the temporal resolution $\Delta t$ of the trajectory, because smaller time steps capture more short-lived fluctuations, leading to an apparent decrease in the computed lifetime.

To reduce statistical noise and assess uncertainties, we divided the trajectory into 50 blocks, each of 200 time steps (corresponding to 2 ps). For each block, $C_c(t)$ was averaged over all hydrogen bonds of a given type and then fitted with a single-exponential decay function:

$$C_c(t) \approx e^{-\frac{t}{\tau_c}} \tag{17}$$

where $\tau_c$ is the characteristic lifetime. The fitting was performed over the decaying region of $C_c(t)$, and the goodness of fit was quantified by the coefficient of determination $R^2$, which consistently exceeded 0.95 (typically around 0.98).

The resulting lifetimes from all 50 blocks were then averaged, and the standard deviation of these values was taken as the statistical uncertainty. Finally, the mean block-averaged $C_c(t)$ curve was again fitted to extract the representative lifetime for each H-bond type.

For the intermittent case, temporary bond breaking is allowed as long as the same hydrogen bond reforms later. The corresponding correlation function is given by:

$$C_i(t) = \frac{\langle h(0)h(t) \rangle}{\langle h(0) \rangle} \tag{18}$$

where $h(t) = 1$ if a given pair of atoms is hydrogen-bonded at time t, and $h(t) = 0$ otherwise. Since $h(t)$ is evaluated independently at each time without enforcing continuity, this definition yields a much slower decay of $C_i(t)$, as it accounts for intermittent reformation events. In this case, the full 100-ps trajectory was divided into 10 blocks of 10 ps each. Due to the slower decay, $C_i(t)$ did not reach zero within the simulation window, making exponential fitting unreliable as the obtained $R^2$ values were too low. Therefore, the characteristic lifetimes were determined by direct integration of the correlation function:

$$\tau_i = \int_0^{t_{max}} C_i(t)\, dt \tag{19}$$

As $C_i(t)$ did not fully decay, the resulting absolute values of $\tau_i$ should be considered only as relative indicators, suitable for comparison between different bond types or surface functionalizations. The uncertainties were evaluated analogously to the continuous case, based on the standard deviation between blockwise lifetimes.

# Results and Discussion

The examples of droplet snapshots and corresponding density profiles are shown in Figure 6.

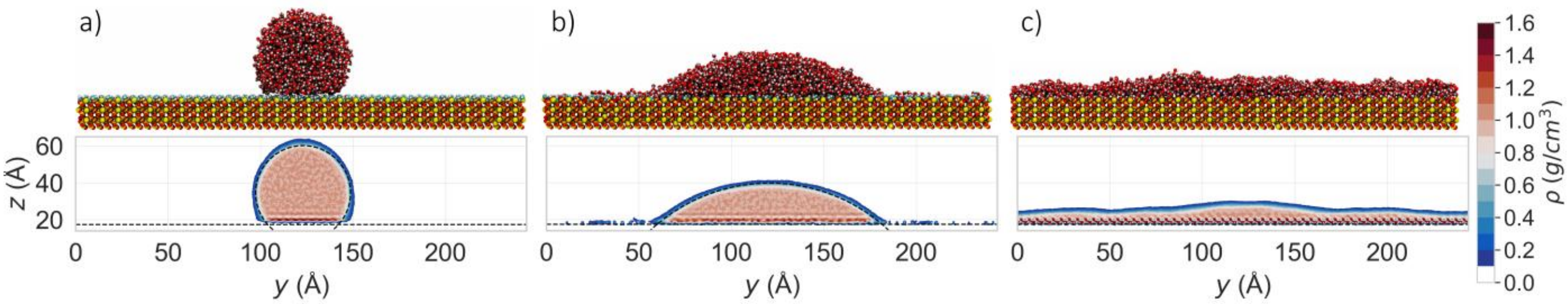

*Figure 6. Examples of droplet snapshots after termalization and averaged density profiles for: a) the most hydrophobic case (methyl functionalization), b) middle case ($N_{OH}/N \cdot 100\% = 50\%$) and c) the most hydrophilic case (hydroxyl functionalization)*

The snapshots depict the droplet morphology, while the density profiles (color gradient) show the variation in mass density, highlighting the transition from a compact droplet to a spread film with increasing hydrophilicity. Additionally, the interfacial water layer becomes increasingly denser.

The resulting temporal dependence of wetting angles for each functionalization is shown in Figure 7a. Variants of 75% and 100% hydroxyl groups functionalization are not shown, because for these cases due to the high hydrophilicity of the surface, the droplet spreads into a film, which aligns with experimental and simulation-based studies at such high hydroxyl concentrations [59,106] . On the opposite end, for the fully methylated surface, our calculated equilibrium contact angle of water is in good agreement with experimental measurements from Sert Çoket et al. [26], who measured the contact angle equal to approximately 140°. These two benchmarks for the limiting surface chemistries demonstrate that our MD-derived contact angles capture both the complete wetting and the hydrophobic regimes observed in real systems.

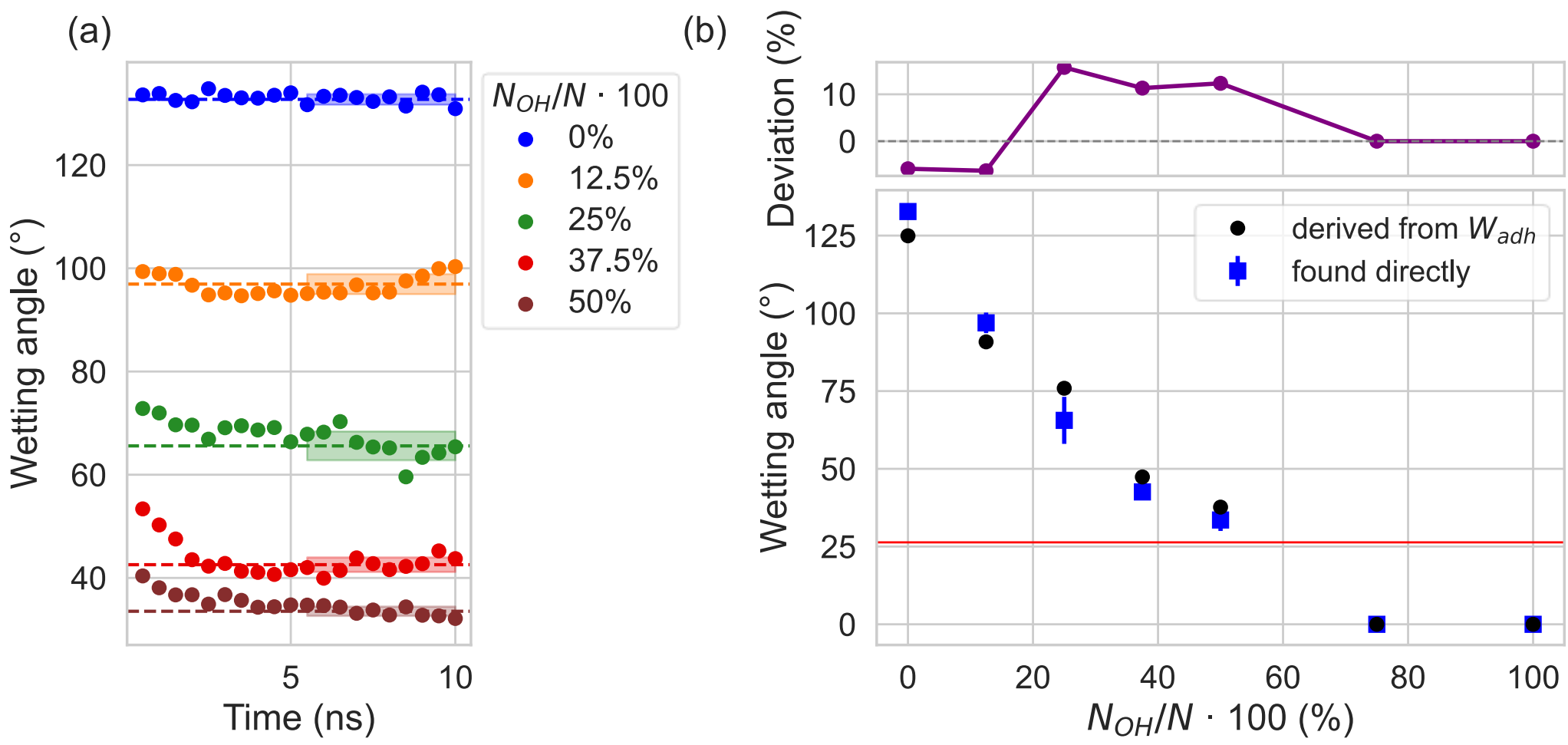

*Figure 7. (a) Time dependence of the wetting angle for the considered variants of surface functionalization. Colored bands highlight the data points used for averaging, while dashed lines and band widths indicate the mean wetting angle and its standard deviation (b) The lower part shows the wetting angle dependence on functionalization degree, calculated directly (blue squares) and derived from the Young-Dupré equation (black circles). The upper part shows the percent deviation of adhesion-derived values from directly calculated values.*

The colored regions in Figure 7a indicate the time intervals used for averaging, while the dashed lines and shaded bands correspond to the mean wetting angle and their standard deviations, respectively. As shown, the contact angle stabilizes after approximately 2–3 ns in all cases, confirming that equilibrium wetting is reached within the simulation time. This behavior is also illustrated in the representative droplet snapshots provided in Figure S1 (for the 50% hydroxylated surface).

The lower part of Figure 7b displays the equilibrium contact angle as a function of the degree of surface functionalization. Blue squares represent contact angles measured directly from the droplet shape, while black circles correspond to values obtained via the Young–Dupré equation[107,108], using the calculated work of adhesion:

$$\theta_{YD} = \arccos\left(\frac{W_{adh}}{\gamma_l} - 1\right) \qquad (20)$$

Here $\gamma_l$ was taken as 63.6 mN/m as was previously reported for the SPC/E water [109]. The upper part of Figure 7b shows the percent deviation of equilibrium contact angles derived from the Young-Dupré equation compared to those calculated directly. The two methods agree well, with a discrepancy of 10–15% for some points. Importantly, this agreement provides a direct validation of the phantom-wall thermodynamic-integration method for the present system, as the work-of-adhesion values obtained from the phantom-wall protocol reproduce independently measured equilibrium contact angles via the Young–Dupré relation. The wetting angle decreases non-linearly with increase in hydroxyl portion. The red line in Figure 7b indicates the wetting angle at which a droplet of a given volume will spread sufficiently so that the distance between it and its periodic image is less than the cut-off distance of the Lennard-Jones potential. Below this angle, interactions between the droplet and its periodic replicas may arise, leading to finite-size artifacts. Therefore, this line represents the lowest reliable contact angle measurable for the given system size.

To confirm that the absence of a resolvable contact angle at high hydroxylation is not caused by limited box size, we repeated the 75% hydroxylation case with the simulation domain doubled along the $y$-axis (see Figure S3 in the SM). The droplet ultimately spread into a thin film, indicating complete wetting. However, this transition took over 35 ns due to strong contact line pinning[20,110] – i.e., the temporary immobilization of the three-phase (solid–liquid–vapor) contact line by local energy barriers and surface heterogeneities, which slowed down the relaxation process. These results confirm that the unresolved contact angle reflects genuine hydrophilic behavior. Accordingly, the work of adhesion is a preferable parameter for characterizing the interaction between a solid and a liquid due to its possibility of application to a wider range of interactions.

Simulation results for the work of adhesion are presented on Figure 8, together with the solid–liquid interaction energy and the entropy gain.

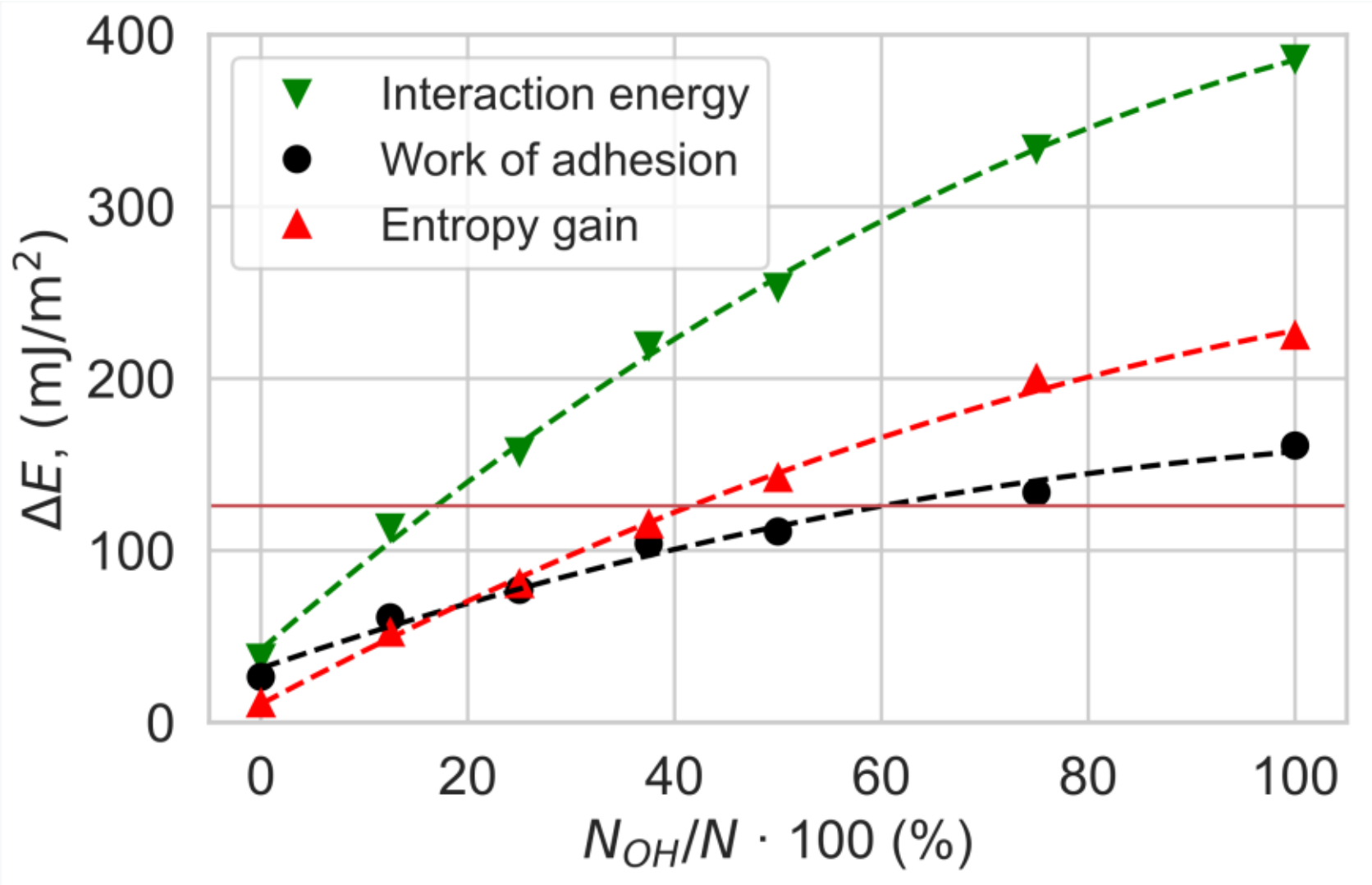

*Figure 8. The dependence the work of adhesion, solid-liquid interaction energy and entropy gain per unit area on the functionalization. Dashed lines serve as visual aids to highlight the trend.*

The results presented in Figure 8 reveal a clear increasing trend in the solid–liquid interaction energy, the work of adhesion, and the interfacial entropy gain as the density of hydroxyl functional groups rises. This behavior is consistent with the enhanced ordering of interfacial water on more hydrophilic surfaces (see Figure 10b and Figure 18). A higher density of hydroxyl functional groups promotes the formation of additional hydrogen bonds (H-bonds) between surface –OH groups and water molecules, as evidenced by the higher H-bonds counts (see Figure 12) for surfaces with greater -OH density. Each H-bond contributes an energetically favorable stabilization (~4–6 kcal/mol), collectively resulting in a more negative solid–liquid interaction energy during interface formation [111]. This H-bonds network enforces greater orientational and positional ordering of interfacial water, reducing its configurational freedom relative to bulk and leading to a pronounced entropy decrease [112]. Consequently, detachment from more hydroxylated surfaces involves breaking a denser H-bond network, requiring a larger enthalpic input ($\varDelta U > 0$). At the same time, the release of strongly ordered interfacial water into the bulk increases entropy ($T\varDelta S > 0$), partially compensating the enthalpic cost. However, the enthalpic contribution dominates, so the total work of adhesion ($W_{adh} = \varDelta U - T\varDelta S$) still increases with higher –OH density. The red line in Figure 8 marks the critical adhesion energy beyond which the droplet no longer maintains a finite contact angle and instead spreads into a thin film. The 75% and 100% hydroxylated cases clearly surpass this limit, indicating that a well-defined contact angle cannot be assigned under these conditions.

The heat transfer between the solid and liquid phases is affected by the effective contact between the atoms of the solid and the liquid, which can be characterized by the magnitude of the depletion length[52]:

$$\delta = \int_0^\infty \left[1 - \frac{\rho_s(z)}{\rho_s^b} - \frac{\rho_l(z)}{\rho_l^b}\right] dz \qquad (21)$$

Here $\rho_s(z)$, $\rho_l(z)$ – are the densities of the solid and the liquid phases, $\rho_s^b$, $\rho_l^b$ – are corresponding densities in bulk region. The depletion length reflects how much the density of the liquid and solid phases deviate from their bulk values at the interface, thus quantifying the extent of the interfacial "gap" where atomic overlap is reduced. Smaller $\delta$ values indicate tighter liquid layering and stronger solid–liquid coupling, which favor higher interfacial thermal conductance. The variation of $\delta$ with surface functionalization is illustrated in Figure 9, where the density distributions for different surface functionalizations are shown: fully methylated, a 1:1 mix of both groups and fully hydroxylated.

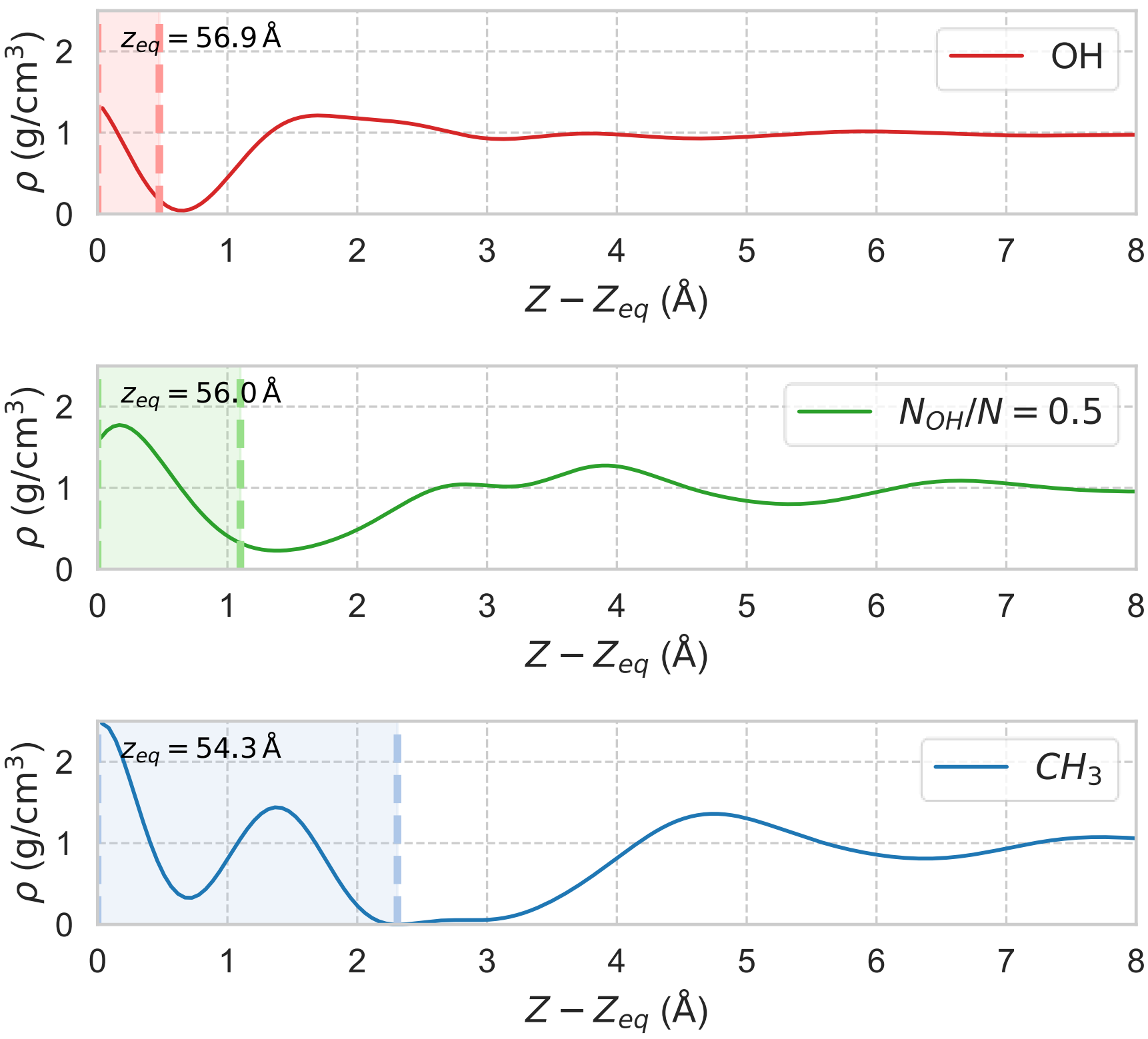

*Figure 9. Density distributions of the liquid phase for three representative surface functionalization variants:* $N_{OH}/N \cdot 100\% =$ *100% (OH), 50%, and 0% (CH3). Each graph is plotted with respect to the equimolar dividing interface* $z_{eq}$*, taken as the origin (i.e., x-axis shows* $z - z_{eq}$ *). The depletion region highlighted by shading.*

These cases are chosen to visually demonstrate the depletion length variation. As the degree of hydroxylation increases, the depletion region becomes narrower. This indicates stronger interfacial structuring and enhanced surface–liquid contact on more hydrophilic surfaces. In particular, the fully hydroxylated surface shows the most pronounced layering near the interface. The full quantitative trend of depletion length across all functionalization levels is presented in Figure 10b. The dependence of ITR on surface functionalization is illustrated in Figure 10a, and its relationship with surface wettability is depicted in Figure 10f.

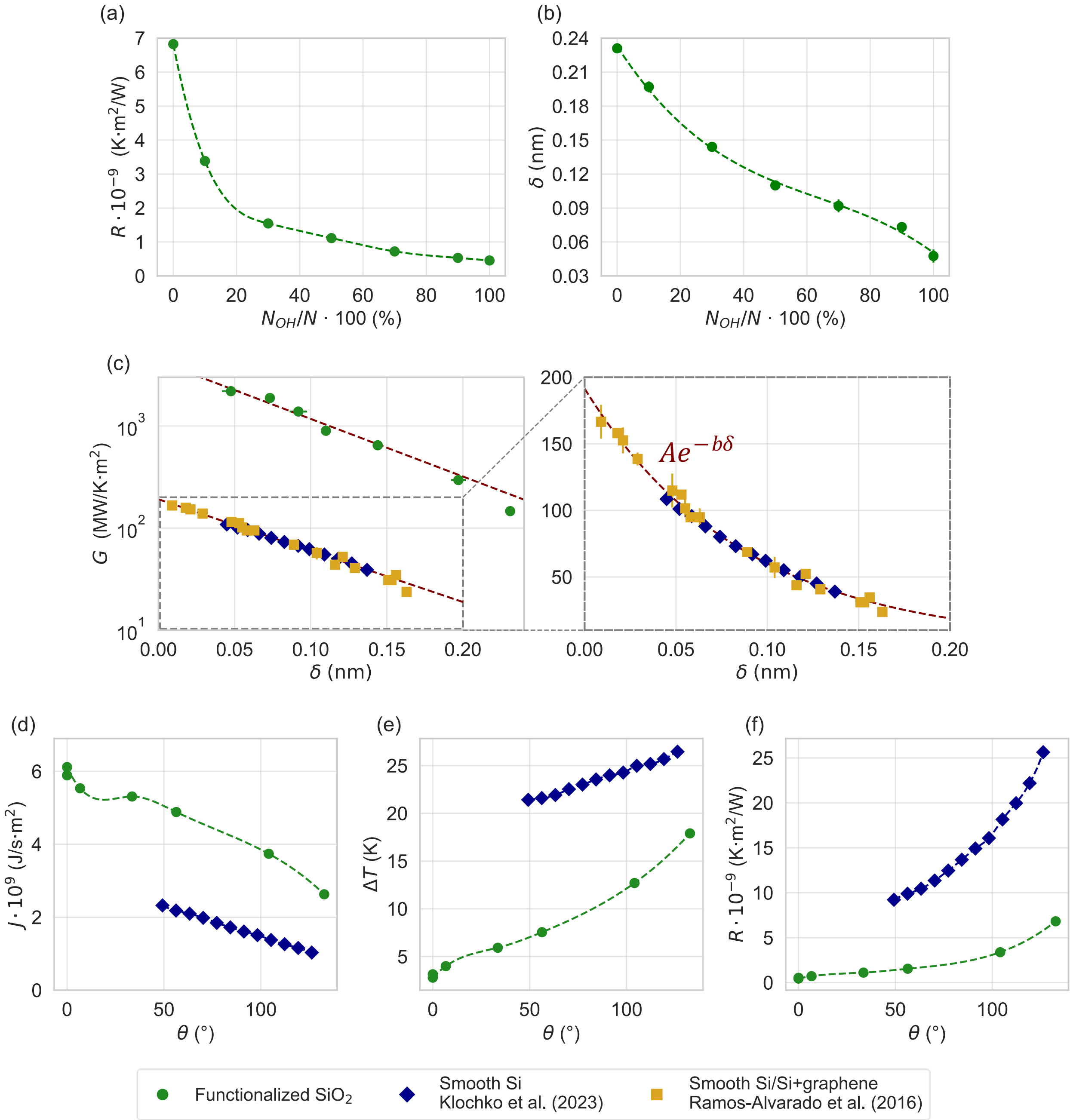

*Figure 10. Dependence of (a) interfacial thermal resistance (ITR) and (b) depletion length on $SiO_2$ surface functionalization (dashed lines are shown as visual guides). (c) Correlation between interfacial thermal conductance and depletion length for functionalized $SiO_2$ (green circles), in comparison with our previous results on smooth Si (100) [45] (blue diamonds) and with literature data for smooth Si (100)/(111) and graphene-coated Si [52] (yellow squares); the dashed line represents a fit. Correlation of (d) heat flux, (e) temperature jump, and (f) thermal resistance with the wetting angle for systems with smooth Si and functionalized $SiO_2$; dashed lines are shown as visual guides. Error bars*

*represent standard deviations over independent NEMD runs (seeds) and are smaller than the marker size.*

Figure 10a reveals a dramatic reduction in thermal resistance with increasing hydroxylation. Even a modest 12.5% –OH coverage results in a twofold drop in ITR, while full hydroxylation decreases the resistance by over an order of magnitude compared to the fully methylated surface. This striking nonlinearity underscores the dominant role of interfacial chemistry in modulating heat transfer.

The absolute values of ITR obtained in this work are of the same order of magnitude as those previously reported for water interfaces with both functionalized amorphous silica[22] and crystalline silica surfaces[23,60]. These earlier studies employed different interatomic potentials and water models, which supports the consistency of the present results with the literature.

While this study focuses on functionalized silica/water interfaces, it is instructive to compare these findings with prior interfacial heat transfer studies involving different material systems. In several works on SAM-functionalized Au substrates, a consistent increase in interfacial thermal conductance is reported upon replacing hydrophobic groups (–$CH_3$, –$CF_3$) with hydrophilic ones (–OH, –COOH), with 2–3-fold enhancements observed depending on the specific end group and bonding configuration [46,49]. However, the absolute conductance values in such systems remain lower than those reported for hydroxylated silica. In contrast, a silica/SAMs/water system investigated by Goicochea et al. [15] exhibited even higher conductance than fully hydroxylated silica, attributed to covalent bonding and strong hydrogen-bonded SAM–water interactions. The MD study by Gonçalves & Termentzidis[22], which directly examined amorphous functionalized silica, showed a ~4.5× increase in interfacial resistance when silanol groups were replaced with hydrophobic trimethylsilanes. Sun et al. [113] reported a similar ~2-fold ITC gain (ITR decrease) for ionized silica (0–50% $SiO^-$ $Na^+$), attributing it to denser interfacial water and more H-bonds, though their uniform silanol baseline (4.7 OH/$nm^2$) yielded higher baseline conductance (~750 MW/$m^2$K) than our fully methylated case (~120 MW/$m^2$K). For H-bonding organics, Sun et al. [114] found ITC rising nonlinearly with silanol density (peaking at ~6.9 OH/$nm^2$ before plateauing due to silanol-silanol competition), similar to our nonlinear ITR drop that accelerates for small hydroxyl density range and then moderates, validating the effective hydroxyl density as a predictor across liquids. Unlike their organic chains, water's smaller size enables tighter layering (depletion δ ~0.3 Å at full OH vs. ~1 Å in[114]), amplifying conductance.

Similar to the ITR dependence, the depletion length also decreases with increasing hydroxyl concentration, though not as drastically. In order to compare the results with the existing ones we also calculated the depletion length for the systems examined in our previous work [45], where we analyzed smooth silicon surfaces with varying wettability. Additionally, we compared our results with those of Ramos-Alvarado et al. [52], who investigated silicon with different crystal planes and silicon coated with graphene. The results are presented on Figure 10c. Here the brown lines are approximating lines of function $Ae^{-b\delta}$ with the parameters: $A = 4302.94$ MW/Km$^2$ and $b = 12.99$ nm$^{-1}$ for silica and $A = 191.48$ MW/Km$^2$ and $b = 11.58$ nm$^{-1}$ for silicon.

From Figure 10c it is evident that the interfacial thermal conductance can differ by up to an order of magnitude between different surfaces with the same depletion length. This indicates that the layering of interfacial water alone, quantified here by the extent of density depletion, is insufficient to predict the thermal boundary conductance. Additional factors such as the atomic-scale structure of the solid surface and the orientation of liquid molecules near the surface may contribute significantly to heat transport.

To further explore this observation, we compared the dependence of temperature jump, heat flux, and ITR on the wetting angle for smooth silicon surfaces (from our previous study [45]) and for the functionalized silica surfaces investigated in the current work. As shown in Figure 10d–f, for similar wettability values functionalized silica exhibits a markedly smaller temperature jump and higher interfacial heat flux compared to silicon, resulting in a significantly lower interfacial thermal resistance. In some cases, the difference reaches nearly an order of magnitude. These findings indicate that interfacial heat transport is not governed by wettability or depletion length alone, but also by the specific nature of interfacial interactions.

To investigate in more detail how functional groups affect heat transfer, we divided the heat flux into parts depending on which group contributes to it (see Equation ( 7 )). Figure 11 shows the share of heat flux in percent that is accounted for by bulk atoms, hydroxyl and methyl groups for each considered functionalization variant.

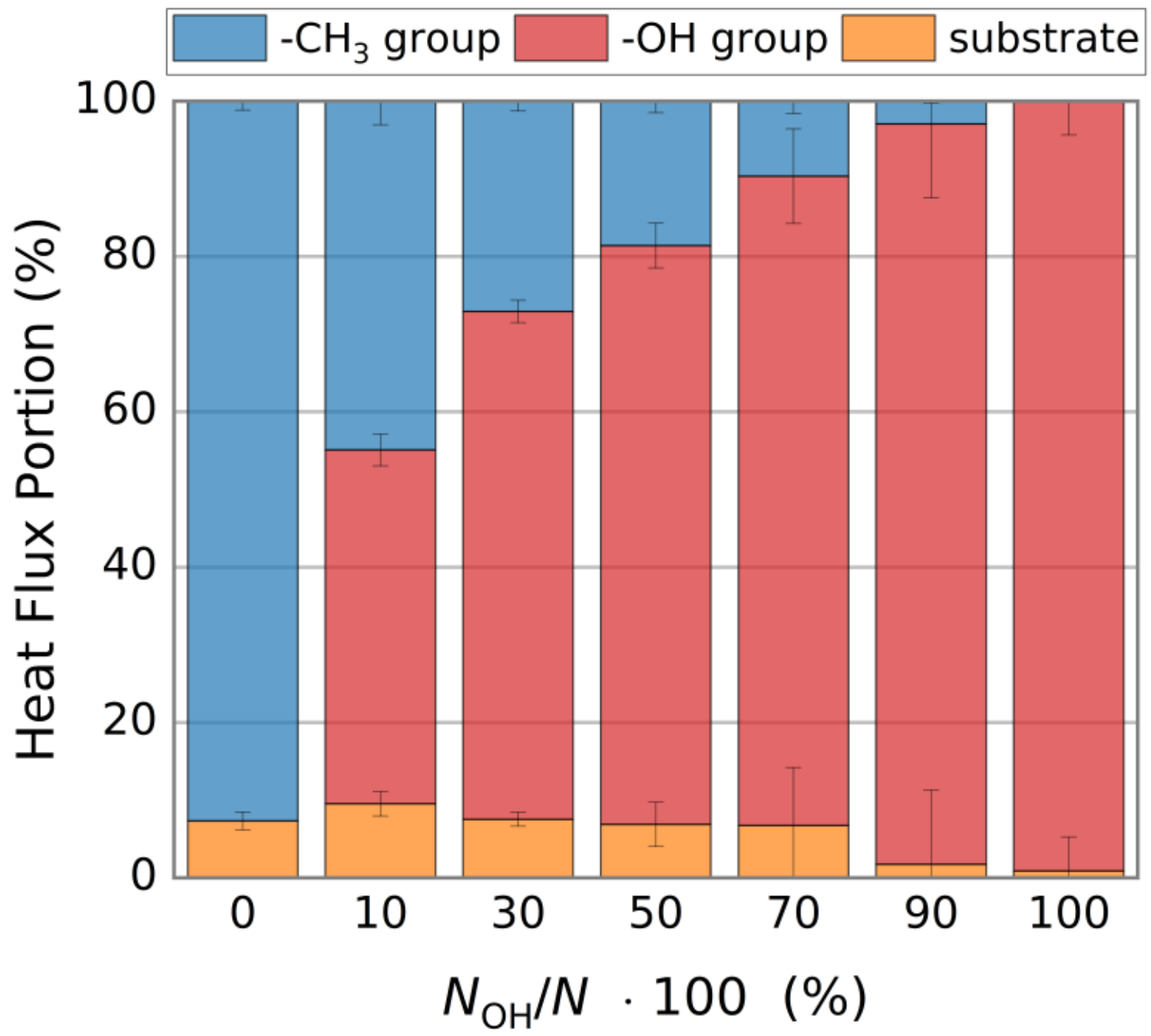

*Figure 11. The contribution of the different functionalized groups and the substrate to the total heat flux*

Figure 11 illustrates that for the fully methylated surface ($N_{OH}/N \cdot 100\% = 0\%$), the substrate contributes approximately 8% to the total interfacial heat transfer. This may initially appear counterintuitive, given the weak interaction between methyl groups and water, and considering that the ClayFF potential tends to overestimate silica–water interactions. However, due to the hydrophobic nature of the methyl-functionalized surface, the water molecules reside at a greater distance from the underlying substrate. As a result, thermal coupling through the substrate is limited, and most of the heat is transferred directly via the methyl groups despite their weaker coupling efficiency.

Interestingly, the substrate's contribution remains relatively constant up to ~90% hydroxylation, suggesting that in these systems, the heat transferred through the substrate is mainly associated with regions beneath methyl-covered sites. This is confirmed by the fully hydroxylated surface ($N_{OH}/N \cdot 100\% = 100\%$), where the substrate accounts for less than 1% of the total heat flux. In this case, the strong hydrogen bonding between –OH groups and water dominates the interfacial coupling, effectively "screening" the substrate from the liquid phase.

Notably, the case with 10% hydroxylation exhibits a slightly higher substrate contribution than the fully methylated case. This may also seem counterintuitive, as the overall area covered by methyl groups is reduced, and previous observations suggest that substrate-mediated heat transfer primarily occurs through regions beneath methyl-functionalized sites. However, the introduction of a small number of hydroxyl groups changes the interfacial configuration: these groups attract water molecules closer to the surface, enabling them to interact not only with the –OH sites but also with the underlying silica that was previously shielded by methyl groups. As a result, although the methyl-covered area decreases, the local proximity of water to the substrate increases elsewhere, partially compensating, and in this case, even enhancing, the substrate's overall contribution. This explains why the substrate heat flux fraction does not decrease proportionally with methyl coverage.

Another noteworthy result is the comparison of the contributions from the functional groups themselves. In the case with only 10% surface coverage by hydroxyl groups, these –OH groups still account for approximately 50% of the total heat transfer. This disproportionate contribution highlights their exceptional efficiency in mediating interfacial thermal transport.

To better understand the microscopic origin of this behavior, we next examine average number of hydrogen bonds formed by the functional groups on the surface. The average number of hydrogen bonds, categorized as $CH_3$ donor, OH donor, and OH acceptor, is shown in Figure 12. To improve clarity, schematic representations of each H-bond type are shown alongside the bar chart.

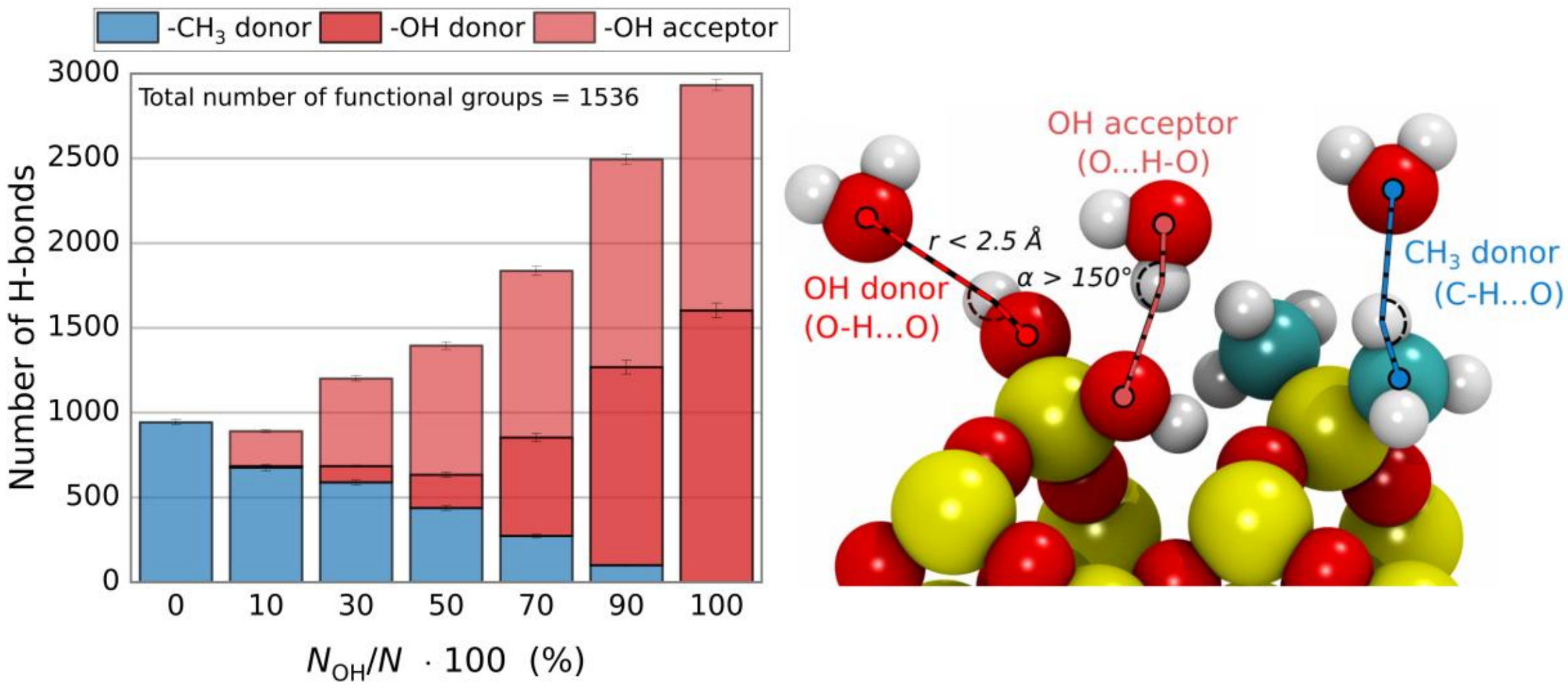

*Figure 12. Left: Average number of hydrogen bonds formed by surface functional groups as a function of hydroxylation percentage. Blue bars represent $CH_3$ donor H-bonds, red bars represent OH donor H-bonds, and light red bars represent OH acceptor H-bonds. Right: Representative snapshot showing examples of each H-bond type: $CH_3$ donor (C–H…O), OH donor (O–H…O), and OH acceptor (O…H–O)*

The analysis reveals distinct trends in hydrogen bonding with varying surface hydroxylation. At 0% hydroxylation (fully methylated surface), only $CH_3$ donor H-bonds are present, with an average of near 950 H-bonds (while the total number of $CH_3$ groups is 1536). The comparatively moderate number of $CH_3$ donor hydrogen bonds can be explained by the molecular arrangement observed in thermalized surface snapshots (see Figure 1). Due to steric effects, one of the two $CH_3$ groups bound to each surface silicon atom often points inward, toward the substrate, and is thus not accessible to the water phase for hydrogen bonding. As the surface becomes increasingly hydroxylated, the number of $CH_3$ donor H-bonds decreases proportionally, while both OH donor and OH acceptor H-bonds increase significantly. Notably, even at low hydroxylation levels, such as 10%, the OH groups are responsible for a substantial number of H-bonds, particularly as acceptors (205 OH acceptor H-bonds compared to 9 OH donor H-bonds). This predominance of OH acceptor bonds at low hydroxylation can be explained by the local steric and electrostatic environment created by neighboring $CH_3$ groups. The hydrophobic methyl groups exert a lateral pressure on isolated hydroxyls, forcing their O–H vectors to orient toward the silica substrate rather than toward the interfacial water [115]. As a result, such OH groups are hindered from donating hydrogen bonds but remain accessible as strong acceptors through their surface oxygen atoms. The sharp increase in H-bonds formed with OH bonds share aligns with the observation in Figure 11 that OH groups account for a disproportionate share of the heat flux at low coverage (e.g., 50% of heat transfer at 10% hydroxylation). Nevertheless, although hydroxyl groups form about three times fewer hydrogen bonds than methyl groups, their stronger hydrogen bonding allows each OH group to transfer more heat per bond, resulting in an overall heat transfer contribution across the interface comparable to that of the more numerous $CH_3$–mediated interactions. This is explained by the difference in the nature of these bonds. Specifically, the hydrogen bonds formed by methyl groups ($CH_3$ donor H-bonds) are generally weaker and less energetic compared to those formed by hydroxyl groups (OH donor and acceptor H-bonds)[100]. This difference arises because the hydrogen atoms in methyl groups are less polar due to the lower electronegativity of carbon compared to oxygen, resulting in a weaker partial positive charge on the hydrogen atom. As a result, the C-H…O hydrogen bonds are less stable and have lower bond energies (typically 1–3 kcal/mol) compared to the O-H…O hydrogen bonds (4–6 kcal/mol)[100]. At 100% hydroxylation, the total number of hydrogen bonds reaches its peak, with OH donor bonds dominating (∼1600) compared to ∼1330 OH acceptor bonds. Given that there are 1536 hydroxyl groups in total, almost each group forms two H-bonds on average.

Figure 13 presents the average rate of interfacial H-bond formation for each functionalization variant, normalized by the total count of corresponding surface groups (–OH or –CH3). The y-axis indicates the number of newly formed or broken H-bonds per functional group per picosecond, i.e. $N_{H-bonds}/(N_{groups} \cdot \Delta t)$ . The corresponding rupture rates were found to be almost identical, with the formation-to-rupture ratio remaining within 0.05 % of unity for all cases. This indicates that all simulated systems reached a stable dynamic equilibrium, with no net gain or loss of hydrogen bonds over time.

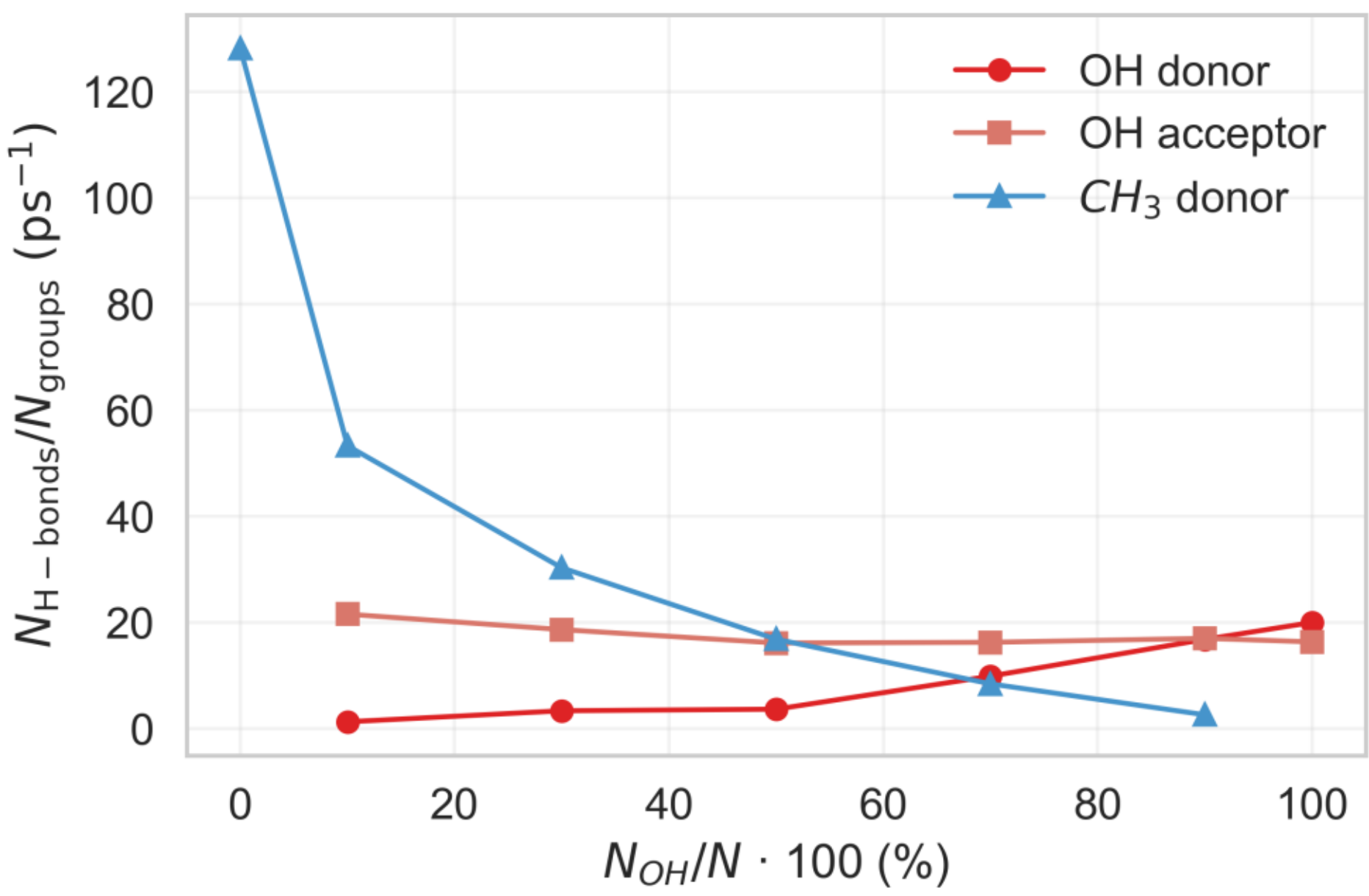

*Figure 13. Hydrogen-bond formation/rupture rate per surface functional group as a function of hydroxylation percentage.*

The data reveal a pronounced dependence on surface chemistry. As the surface becomes increasingly hydroxylated, the formation rate of $CH_3$–donor hydrogen bonds decreases sharply. Specifically, replacing only 10 % of the $CH_3$ groups with –OH groups reduces the $CH_3$–donor formation rate by approximately a factor of two. This strong decline correlates with the corresponding decrease in the fraction of heat flux carried through methyl sites (see Figure 11), supporting the close link between hydrogen-bonding activity and interfacial heat transport efficiency.

In contrast, the formation rate for OH– acceptor bonds remains nearly constant across the entire range of hydroxylation levels, and it consistently exceeds that of OH–donor bonds. This behavior suggests that the oxygen atoms of surface hydroxyls maintain a high capacity to attract hydrogen from water molecules, even when their surface density changes. Meanwhile, the OH donor formation rate increases progressively with surface hydroxylation, surpassing that of acceptors at full hydroxyl coverage. This trend is consistent with the dominance of OH donor bonds observed in the fully hydroxylated system (Figure 11) and further supports the notion that these strong O–H···O interactions are the principal mediators of interfacial energy exchange.

Figure 14 presents the decay of the hydrogen bond autocorrelation functions computed using both continuous (solid line) and intermittent (dashed line) definitions for the functionalization variant of 50% hydroxylation:

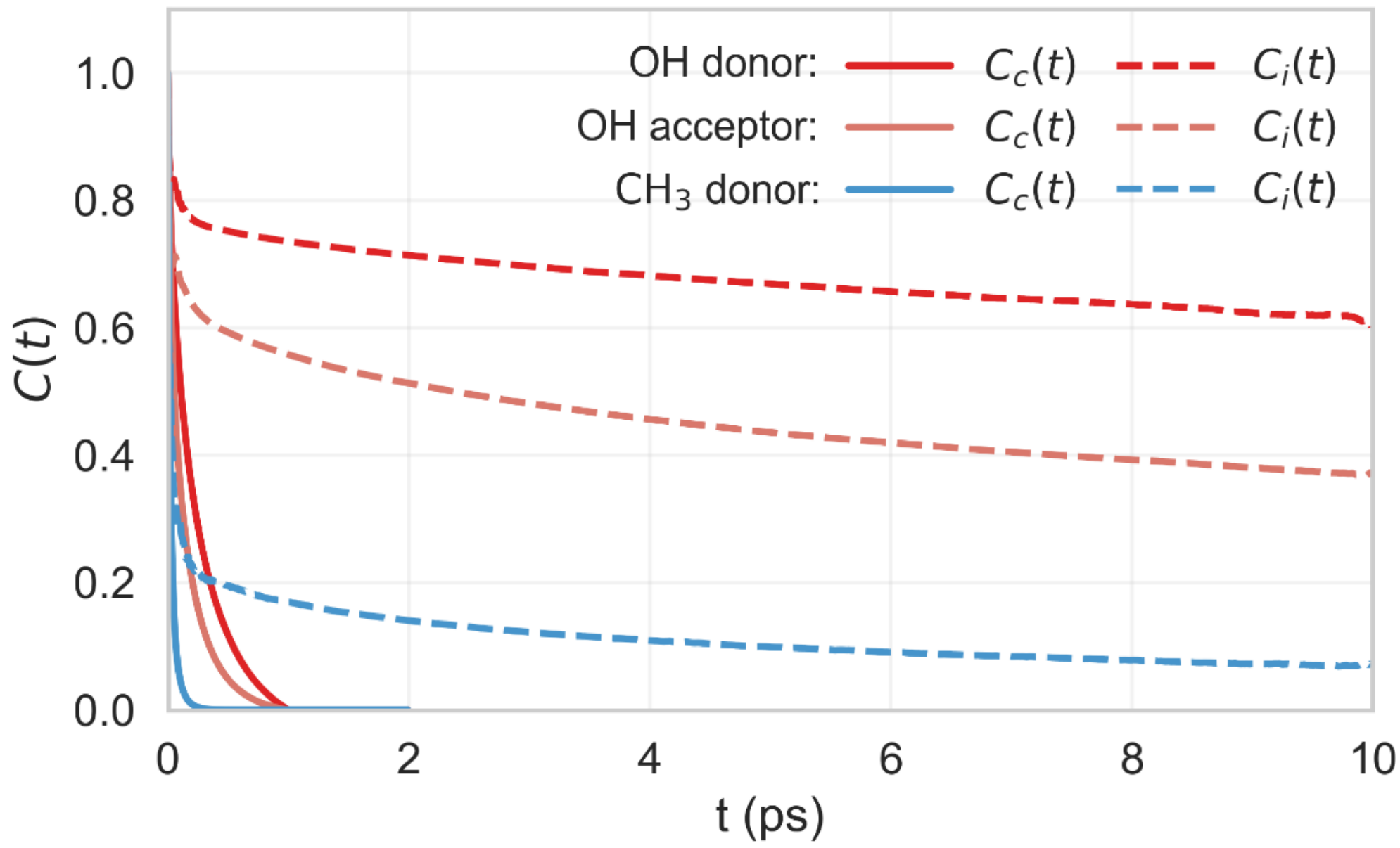

*Figure 14. Hydrogen bond autocorrelation functions for the functionalization variant $N_{OH}/N \cdot 100\%$ = 50% computed using continuous (solid lines) and intermittent (dashed lines) methods.*

As can be seen, the two approaches yield noticeably different decay behaviors. The continuous definition produces a much faster decay comparing to the intermittent definition. Nevertheless, both methods reveal the same relative trend among the different types of hydrogen bonds: The $CH_3$ donor bonds decay the fastest, indicating their lowest stability, while the OH donor bonds exhibit the slowest decay and thus the highest persistence. The OH acceptor bonds fall in between these two limits.

This consistent hierarchy across both correlation definitions confirms that the stability ordering is intrinsic to the chemical nature of each bonding group rather than an artifact of the chosen analysis method.

Figure 15 presents the variation of the hydrogen bond lifetimes as a function of the surface hydroxylation degree.

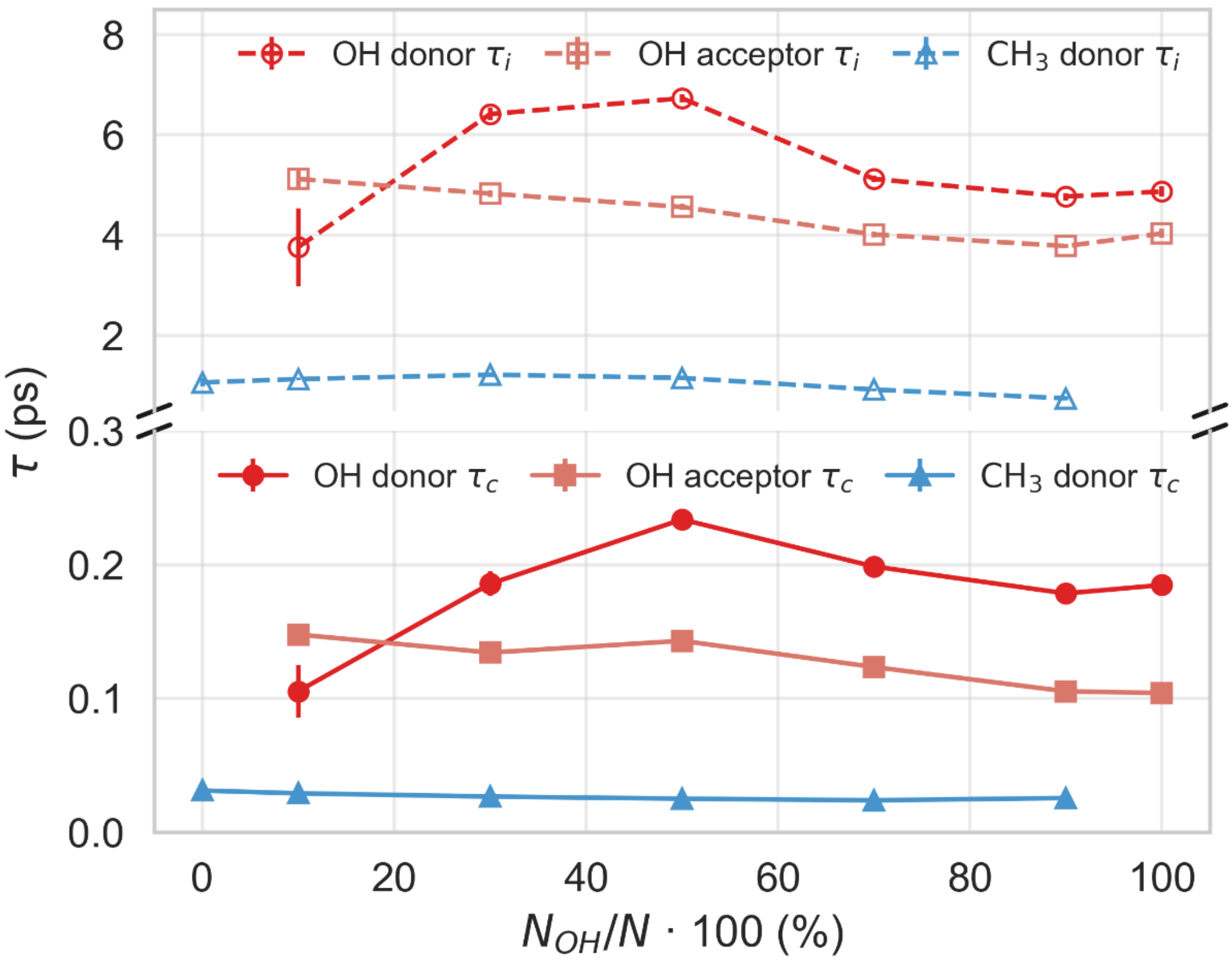

*Figure 15. Average lifetimes of hydrogen bonds as a function of surface hydroxylation degree. Solid symbols correspond to the continuous definition ($\tau_c$), and open symbols represent the intermittent definition ($\tau_i$). Error bars indicate the standard error of the mean.*

The overall trends obtained from the two methods are consistent with each other: despite the quantitative difference in the absolute magnitudes of $\tau_c$ and $\tau_i$, both show the same relative ordering among the different types of hydrogen bonds.

The $CH_3$ donor bonds exhibit the shortest lifetimes, about five to six times smaller than those of OH donor bonds, confirming their weak and highly dynamic nature. Among the hydroxyl-related bonds, OH donors systematically show longer lifetimes than OH acceptors by roughly a factor of 1.5, except at 10% hydroxylation where the opposite trend is observed. This inversion can be attributed to the local structural effects induced by the surrounding methyl groups: at low hydroxylation levels, isolated OH groups are sterically confined and tend to orient their hydrogen atoms toward the silica substrate rather than the liquid phase, as evidenced by the predominance of OH acceptor bonds observed in Figure 12. Such an orientation suppresses the formation of stable OH donor interactions, making the few donor bonds that do form both rare and short-lived.

As the hydroxylation degree increases, the lifetime of OH donor bonds first increases markedly up to ~50% coverage, as –OH groups gain orientational freedom, allowing their hydrogens to point toward the liquid phase and participate in more persistent donor H-bonds. However, at higher hydroxylation levels, the dense distribution of –OH groups may facilitate frequent H-bond switching and the formation of transient inter-surface OH···OH interactions [114]. These interactions compete with water for both hydrogen and oxygen sites: donor bonds become less stable as surface hydroxyls intermittently bond with neighboring groups, while acceptor sites become partially saturated by such internal interactions. As a result, the lifetime of both donor and acceptor H-bonds with water decreases beyond ~50% coverage.

The strong hydrogen bonding capability of OH groups likely enhances the thermal coupling between the surface and water, supporting our assertion that hydrogen bonds favor interfacial heat transfer. The dependence of ITR on the number of interfacial H-bonds is presented in Figure 16.

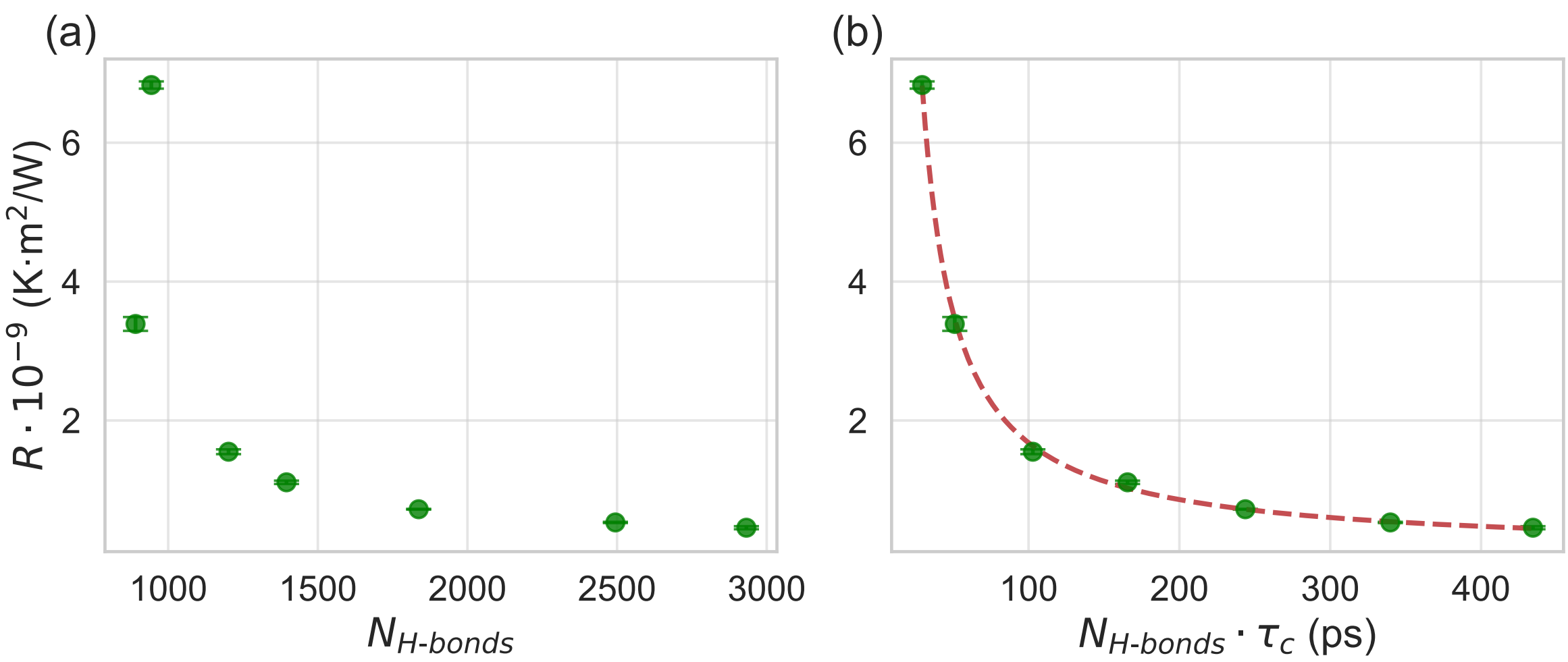

*Figure 16. ITR as a function of hydrogen bonding at the interface: (a) versus the average number of surface–water hydrogen bonds ($N_{H-bonds}$). (b) versus the effective H-bond metric ($N_{H-bonds} \cdot \tau_C$) where each contribution is weighted by the corresponding H-bond lifetime*

A clear nonlinear decrease in ITR is observed with increasing number of interfacial H-bonds ($N_{H-bonds}$), with the improvement becoming progressively weaker at high $N_{H-bonds}$. A similar behaviour was reported by Sun et al. [114], where silanol-silanol interactions saturated available sites and limited liquid-mediated bonding, leading to a plateau in ITC. Besides, the point in Figure 16a corresponding to the highest resistance deviates from this trend. This suggests that counting H-bonds alone is insufficient, and that their lifetimes must also be taken into account. When the product $N_{H-bonds} \cdot \tau_C$ is used instead of $N_{H-bonds}$, the correlation becomes smooth across all coverages (Figure 16b), indicating that the effective connectivity of the interfacial H-bond network is the relevant quantity controlling thermal transport.

After analyzing the contribution of interfacial groups and the role of hydrogen bonding in heat transfer, we now turn our attention to the structure of the adjacent water layer itself. Specifically, we examine the orientation of interfacial water molecules, which provides insight into how surface chemistry shapes interfacial hydrogen bond networks [98], as this organization can also play a role in mediating thermal transport [23,116,117]. Figure 17 shows the probability distribution of orientations of a water molecule relative to the substrate surface.

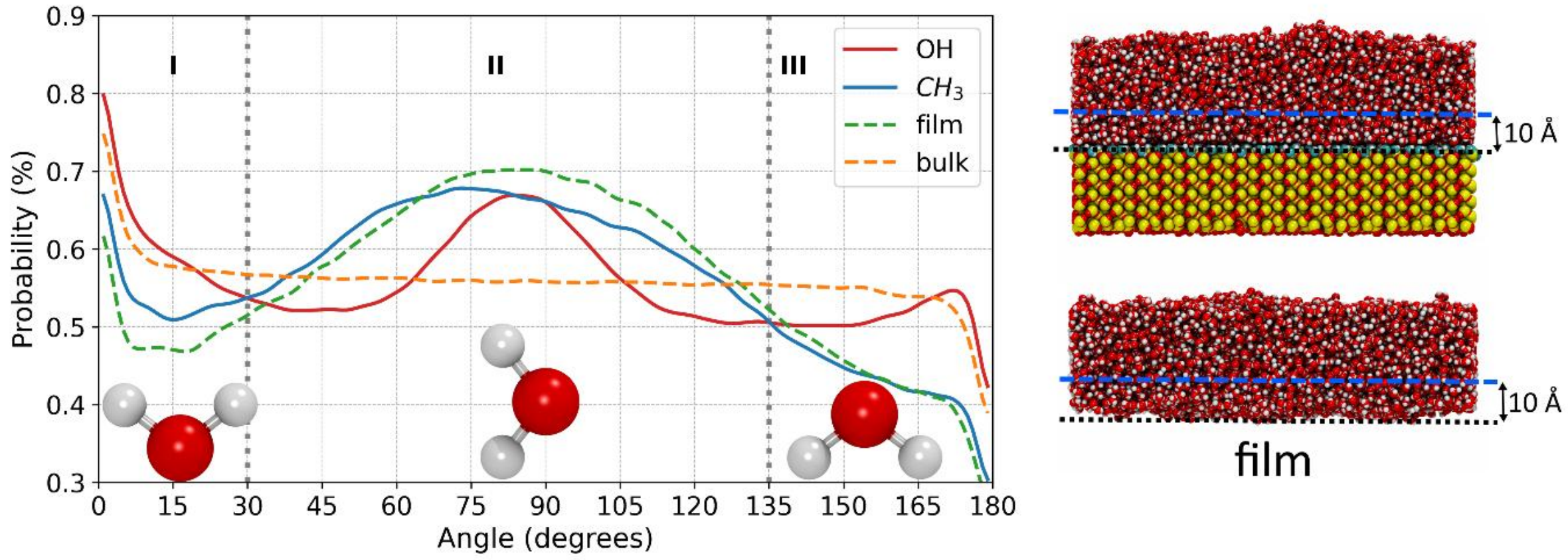

*Figure 17. Probability distribution of water molecule orientations relative to the surface normal. The red and blue curves show the distributions for OH- and $CH_3$-functionalized surfaces, respectively. For comparison, the green and orange dashed curves represent the film (water/vacuum interface) and bulk water. The snapshots on the right show the region where water molecules orientations were taken into account.*

The probabilities of water molecule orientations were analyzed in three angular regions $[\theta_a, \theta_b]$: 0–30°, 30–135°, and above 135° denoted as I, II, III. In the case of the hydrophilic (OH-functionalized) surface (red curve) and the hydrophobic ($CH_3$-functionalized) surface (blue curve), notable differences were observed. On the OH-covered silica surface, water molecules exhibit a pronounced tendency to align along the surface normal, with a normalized probability $f_{[\theta_a,\theta_b]}$ = 40.30% in the 0°–30° and 135°–180° ranges, compared to 34.72% for the CH3-covered surface - a relative increase of 16.1%. This aligns with molecular dynamics studies showing that hydroxylated silica surfaces promote ordered water layers with hydrogen-down orientations [113,118]. The pronounced perpendicular orientation on hydrophilic surfaces promotes efficient H-bond thermal energy paths normal to the interface [113,114]. Conversely, on the $CH_3$ surface, water molecules are 9.3% more likely to align along the surface plane (30°–135° range) than on the OH surface, reflecting the hydrophobic nature of $CH_3$ groups. At the water/vacuum interface (film), water molecules show the least tendency to align along the normal, with a probability of lying along the plane higher than both OH and $CH_3$ surfaces. This planar preference arises from the absence of a solid surface to anchor the molecules, allowing surface tension and dipole interactions to dominate. In contrast, the bulk water exhibits a near-isotropic distribution, serving as a baseline for comparison. These findings confirm that water near the functionalized silica surface is oriented differently depending on the surface chemistry, which should have implications for interfacial phenomena.

It is important to note that the observed asymmetry in the distributions stems from the single-interface configuration of our system, where water is present only above the substrate. This results in an asymmetric distribution of orientations, in contrast to the symmetric curves reported by Wang et al. [98] for confined water in nanoporous silicon. The deviation from symmetry is also evident in the film case (see the lower snapshot in Figure 17), where the water/vacuum interface (with vacuum positioned below) yields a similar asymmetric trend. The bulk water distribution is nearly uniform, as expected from the isotropic nature of the system. Additionally, minor anomalies at the angular limits are attributed to the normalization procedure (division by sin(θ), see Equation ( 13 )).

To provide a compact quantitative measure of water molecules orientation differences, we computed the local orientational order parameter $S$ for water molecules in the immediate vicinity of –OH and –$CH_3$ groups. In both cases, the order parameter is negative: $S_{OH} = -0.12$ and $S_{CH_3} = -0.25$ indicating that interfacial water dipoles preferentially adopt orientations closer to the interfacial plane rather than along the surface normal. Nevertheless, the less negative value near –OH groups confirms that hydroxyl terminations promote a greater degree of normal alignment compared to methyl groups, in agreement with the angular probability distributions in Figure 17.

To complement the analysis of orientation, we further examined the connectivity of the hydrogen bond network through geodesic distance analysis in the same interfacial region. The geodesic distance here is defined as the average number of hydrogen bonds forming the shortest path between two water molecules within the largest connected hydrogen-bond cluster, and thus it is a dimensionless measure of hydrogen-bond connectivity. The cluster refers to a group of water molecules connected through at least one hydrogen bond, forming a single continuous network. Surprisingly, the average geodesic length was found to be larger for the hydrophilic surface (24.15) compared to the hydrophobic surface (23.71). This counterintuitive result can be attributed to the interaction between water molecules and the surface. In the hydrophilic case, water molecules tend to form hydrogen bonds with the surface, which can saturate some of their bonding capacity, increasing the geodesic distances. Additionally, the hydrophilic surface exhibits a larger maximum cluster size (2737.6) compared to the hydrophobic surface (2591.7), which can also contribute to longer geodesic paths within the network.

To consolidate our understanding of the interfacial water structure and its interaction with the surface, we calculated the radial distribution function $g(r)$ between the surface and water oxygen atoms. Figure 18 shows the $g(r)$ of water O atoms with O in –OH groups and C in $-CH_3$ groups for two extreme cases.

It should be noted that the radial distribution functions shown here were calculated assuming spherical shells around surface atoms. In our system, the water layer is of finite thickness and sits on a solid surface with vacuum above. Consequently, at larger distances, parts of the spherical shells extend outside the liquid region, which breaks the isotropic assumption used in bulk RDF calculations. As a result, $g(r)$ does not converge to unity but reaches elevated values ($\sim$2.5). This behavior is a well-known and expected feature of RDFs in interfacial or confined systems [119], reflecting the system geometry rather than any lack of equilibration or structural disorder. Importantly, the observed layering and peak structure near the surface remain fully representative of the interfacial water organization.

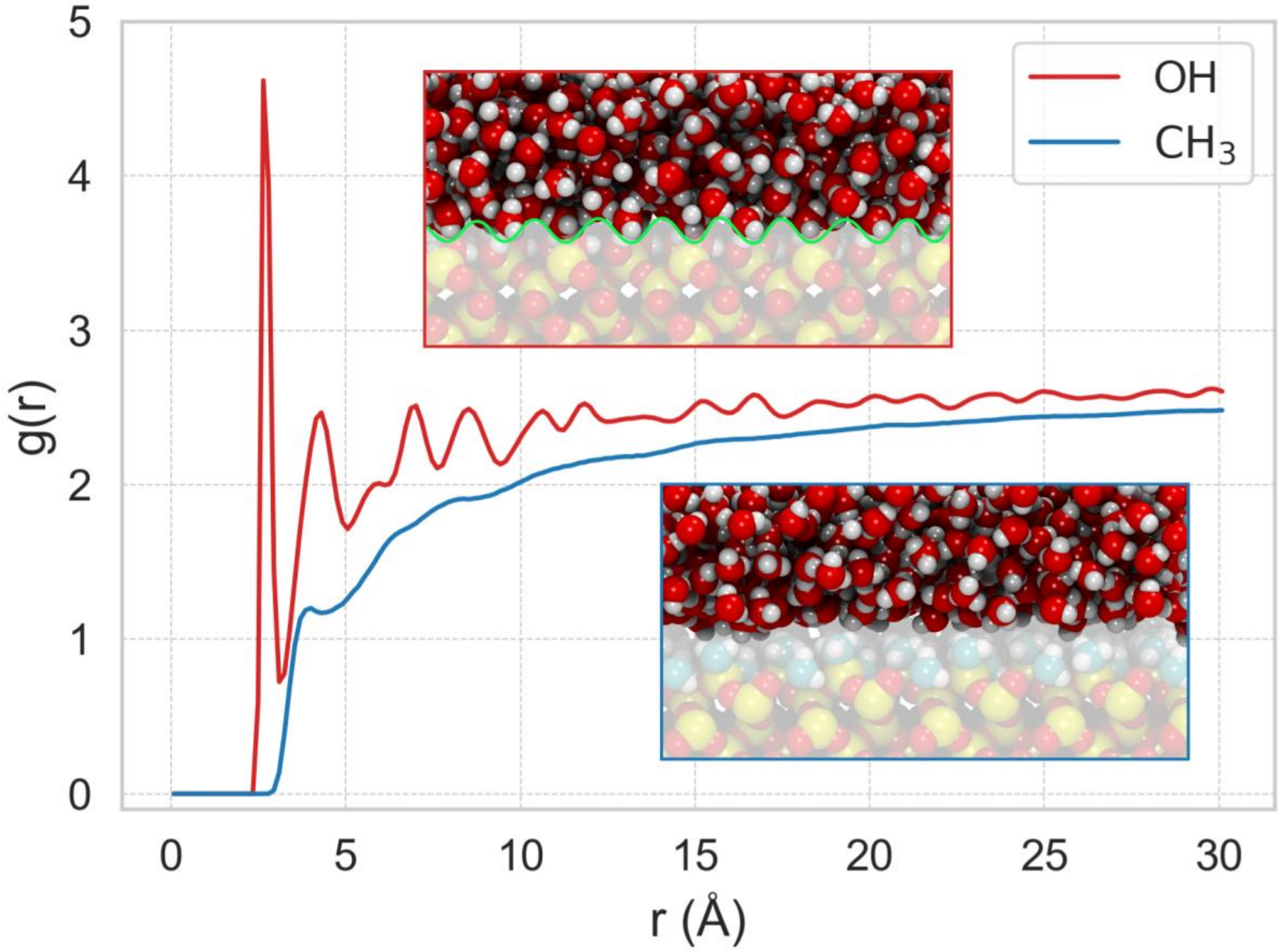

*Figure 18. Radial distribution functions $g(r)$ of surface – water oxygen atoms near silica surfaces functionalized with –OH (red) and –$CH_3$ (blue) groups, and representative snapshots of water structure at the interface. The green wavy line in the snapshot visually marks the approximate front formed by the water molecules.*

In the hydrophilic case, the first $g(r)$ peak appears at 2.64 Å, indicating strong short-range interactions between water molecules and the –OH terminated surface. Multiple subsequent peaks are also visible, reflecting the formation of a layered, ordered water structure due to hydrogen bonding and surface affinity. In contrast, the hydrophobic surface exhibits a much broader and delayed first peak at 4 Å, with no pronounced secondary peaks, suggesting weak interaction with water and the absence of any significant structuring. The contrast between the sharp, layered peaks near –OH groups and the diffuse structure near –$CH_3$ groups directly illustrates the superior ability of hydroxylated surfaces to anchor and organize water, which, as shown earlier, is essential for efficient heat transfer. The difference is seen on the snapshots, where the water near

the –OH surface shows visible structuring, in contrast to the more diffuse distribution near the $–CH_3$ surface. This difference arises from the contrasting chemical nature of the functional groups: –OH groups form hydrogen bonds with water, promoting alignment and layering, while $–CH_3$ groups are nonpolar and repel water molecules, leading to a more disordered, bulk-like distribution.

Thus, despite the larger geodesic distances within water clusters observed in the hydrophilic case, the structured water layer along with perpendicular water molecule orientation likely enhance heat transfer properties, aligning with our initial hypothesis.

# Conclusions

The MD simulations performed in this study confirm that surface functionalization of silica has a profound impact on interfacial water structure and heat transfer. By tuning the ratio of hydroxyl (–OH) to methyl ($–CH_3$) groups, it is possible to effectively control wettability, adhesion energy, and ITR.

Hydroxylation enhances hydrophilicity and enables water molecules to approach the surface more closely, reducing the depletion length and strengthening interfacial bonding. Even a small fraction of OH groups (e.g., 12.5%) can halve the ITR compared to a fully methylated surface, and full hydroxylation yields a tenfold reduction. This striking sensitivity of thermal resistance to chemical composition emphasizes the role of chemical interactions in thermal transport. Water molecules also adopt a more perpendicular orientation near OH-rich surfaces, optimizing contact and energy transfer.

Quantitative analysis of interfacial hydrogen bonds shows that OH groups, although sometimes forming fewer bonds than $CH_3$ groups, contribute disproportionately to heat transfer due to the greater strength and stability of O–H…O interactions. The ITR decreases nonlinearly with the number of interfacial H-bonds, confirming their critical role. Surfaces with only weak $CH_3$ donor bonds exhibit the highest thermal resistance, highlighting the superior coupling efficiency of hydroxyl-driven hydrogen bonding.

The dynamic behavior of interfacial hydrogen bonds was investigated in terms of their formation and rupture rates as well as characteristic lifetimes. The results reveal that the rates and lifetimes are strongly dependent on the surface chemistry. Hydroxylation markedly suppresses the formation of $CH_3$–donor bonds, while simultaneously promoting stronger and more persistent OH–donor interactions with interfacial water. In contrast, OH–acceptor bonds remain less sensitive to surface coverage. The lifetimes obtained from both continuous and intermittent correlation analyses consistently show that $CH_3$–donor bonds are the weakest and most transient, whereas OH–donor bonds exhibit the highest stability, persisting up to five to six times longer. At intermediate hydroxylation levels (~50%), OH donors attain their maximum lifetime, while at higher coverages, the emergence of transient OH⋯OH surface interactions reduces their persistence. This reduction in bond lifetime coincides with the observed slowdown in ITR improvement, suggesting a direct link between the dynamic stability of H-bonds and the efficiency of interfacial thermal transport.

Finally, a comparison with other materials such as smooth silicon and graphene-functionalized surfaces shows that chemical specificity, not just water structuring and surface wettability, are essential in defining interfacial heat transport. Functionalized silica benefits from both optimized structure and chemistry.

These findings provide a comprehensive framework for designing advanced silica-based materials with tailored thermal properties and highlight how atomistic-level understanding can guide the development of more efficient heat exchange systems.

# Supporting Materials

Force-field parameters for bonded and non-bonded interactions used for silica, hydroxyl and methyl surface groups (Tables S1–S2); convergence test of interfacial thermal resistance in NEMD simulations for the $N_{OH}/N \cdot 100\% = 50\%$ hydroxylated surface (Figure S1); time evolution of droplet density maps on the $N_{OH}/N \cdot 100\% = 50\%$ hydroxylated surface (Figure S2); droplet spreading on the enlarged $N_{OH}/N \cdot 100\% = 75\%$ hydroxylated surface (Figure S3); derivation of the discretized formula for the equilibrium interface position between two phases (corresponding to Equation ( *10* )).

# Acknowledgment

This research was supported by the ANR project "PROMENADE" No. ANR-23-CE50-0008. Molecular simulations were conducted using HPC resources from GENCI-TGCC, GENCI-IDRIS and GENCI-CINES (eDARI projects No. A0150913052 and A0170913052). We appreciate Dr Samy Merabia from the Institute of Light and Matter (Institut Lumière Matière, Université Claude Bernard Lyon 1, CNRS).

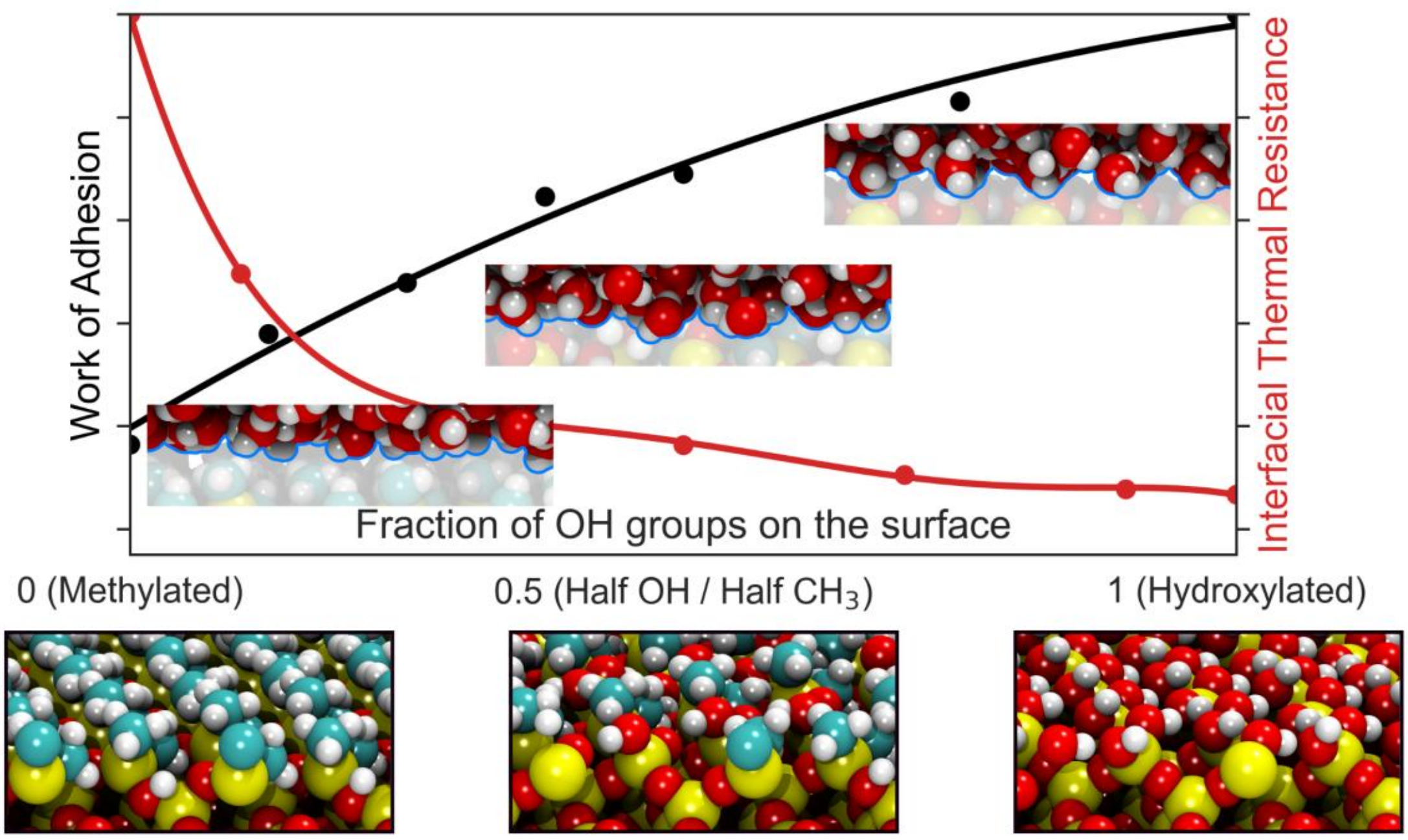

Work of Adhesion
Interfacial Thermal Resistance
Fraction of OH groups on the surface
0 (Methylated)
0.5 (Half OH / Half CH$_3$)
1 (Hydroxylated)

TOC Graphic